# Simple Digital Controls from Approximate Plant Models

Hugh L. Kennedy

*DEWC Services (hugh.kennedy@dewc.com)*

*UniSA STEM Unit, University of South Australia (hugh.kennedy@unisa.edu.au)*

*Mawson Lakes SA 5095*

## Abstract

Two ways of designing low-order discrete-time (i.e. digital) controls for low-order plant (i.e. process) models are considered in this tutorial. The first *polynomial* method finds the controller coefficients that place the poles of the closed-loop feedback system at specified positions for adroit controls, i.e. for a rapid and compressed transient response, when the plant model is known precisely. The poles and zeros of the resulting controller are unconstrainted, although an integrator may be included in the controller structure as a special case to drive steady-state errors towards zero. The second *frequency* method ensures that the feedback system has the desired phase-margin at a specified gain cross-over frequency (for the desired bandwidth) yielding robust stability with respect to plant model uncertainty. The poles of the controller are at specified positions, e.g. for a standard Proportional-Integral (PI), Proportional-Derivative (PD), Proportional-Integral-Derivative (PID), structure or other more general configurations if necessary, and the problem is solved for the controller zeros. The poles and zeros of the resulting closed-loop feedback system are unconstrained. These complementary design procedures allow simple and effective controls to be derived analytically from a plant model, using a matrix inverse operation to solve a small set of linear simultaneous equations, as an alternative to more heuristic (e.g. trial-and-error) or empirical PID-tuning approaches. An azimuth controller for a pan-tilt-zoom camera mount is used as an illustrative example. The ways in which both procedures may be used to design controls with the desired balance between adroitness and robustness are discussed.

# Contents

# Definition of mathematical symbols, operators and acronyms

$i$: Imaginary unit, $i = \sqrt{-1}$.

$\blacksquare^{-1}$: Matrix inverse operation.

$\blacksquare^{T}$: Matrix/vector transpose operator.

$\blacksquare^{*}$: Complex conjugation operator.

$|\blacksquare|$: Magnitude operator, e.g. $|re^{i\varphi}| = r$.

$\angle\blacksquare$: Angle operator, e.g. $\angle re^{i\varphi} = \varphi$.

$\dot{\blacksquare}$: First derivative with respect to time.

$\blacksquare_{M \times N}$ or $[\blacksquare]_{M \times N}$: Dimensions of a matrix. Indicates that the matrix has $M$ rows and $N$ columns.

$(\blacksquare)$: Denotes a function of a continuous argument, e.g. time or frequency.

$[\blacksquare]$: Denotes a vector, or a sampled function in discrete time, that is indexed using an integer argument (from zero).

$F_s$: Sampling rate (in cycles per second or Hertz), i.e. the frame rate of the video camera.

$T_s$: Sampling period (in seconds), $T_s = 1/F_s$.

$T_u$: Time (in seconds) required to compute the control command.

$t$: Time (in seconds).

$n$: Sample index in discrete time, $t = T_s n$.

$m$: Delay index in discrete time, $t = T_s(n - m)$.

dc: Direct Current, i.e. a frequency of zero.

Hz: Hertz, a unit of frequency (in cycles per second).

$F$: Frequency (in samples per second, or Hz).

$f$: Relative frequency (in cycles per sample), $f = F/F_s$.

$\omega$: Relative angular frequency (in radians per sample), $\omega = 2\pi f$.

$\Omega$: Phase-rate coordinate (vertical axis) in the complex s-plane, i.e. angular frequency (in radians per second) $\Omega = 2\pi F$.

$\sigma$: Rate coordinate (horizontal axis) in the complex s-plane (in reciprocal seconds).

$s$: Coordinate in the complex plane reached via the Laplace transform, $s = \sigma + i\Omega$.

$z$: Coordinate in the complex z-plane, reached via the $\mathcal{Z}$-transform.

$\mathcal{L}\{\blacksquare\}$: Laplace-transform operator ($t \to s$).

$\mathcal{L}^{-1}\{\blacksquare\}$: Inverse Laplace-transform operator ($s \to t$).

$\mathcal{F}\{\blacksquare\}$: Fourier-transform operator ($t \to \Omega$).

$\mathcal{F}^{-1}\{\blacksquare\}$: Inverse Fourier-transform operator ($\Omega \to t$).

$\mathcal{Z}\{\blacksquare\}$: $\mathcal{Z}$-transform operator ($n \to z$).

$\mathcal{Z}^{-1}\{\blacksquare\}$: Inverse $\mathcal{Z}$-transform operator ($z \to n$).

$\mathcal{H}(s)$: Transfer function of a continuous-time system, $\mathcal{H}(s) = \mathcal{B}(s)/\mathcal{A}(s)$.

$\mathcal{B}(s)$: Numerator polynomial in $s$.

$\mathcal{A}(s)$: Denominator polynomial in $s$.

$\mathcal{H}(\Omega)$: Frequency response of $\mathcal{H}(s)$, $\mathcal{H}(\Omega) = \mathcal{H}(s)|_{s=i\Omega}$. This simplified notation is used here instead of $\mathcal{H}(i\Omega)$.

$h(t)$: Impulse response of a continuous-time system, $h(t) = \mathcal{L}^{-1}\{\mathcal{H}(s)\}$.

$\mathcal{H}(z)$: Transfer function of a discrete-time system, $\mathcal{H}(z) = \mathcal{B}(z)/\mathcal{A}(z)$.

$\mathcal{B}(z)$: Numerator polynomial in $z$.

$\mathcal{A}(z)$: Denominator polynomial in $z$.

$b[k]$: The $k$th coefficient of the $\mathcal{B}(z)$ polynomial.

$a[k]$: The $k$th coefficient of the $\mathcal{A}(z)$ polynomial.

$\beta_k$: The $k$th root of $\mathcal{B}(z)$, i.e. the $k$th zero of $\mathcal{H}(z)$.

$\alpha_k$: The $k$th root of $\mathcal{A}(z)$, i.e. the $k$th pole of $\mathcal{H}(z)$.

$\gamma$: A gain factor (real).

$h[n]$: Impulse response of a discrete-time system, $h[n] = \mathcal{Z}^{-1}\{\mathcal{H}(z)\}$.

$\mathcal{H}(\omega)$: Frequency response of $\mathcal{H}(z)$, $\mathcal{H}(\omega) = \mathcal{H}(z)|_{z=e^{i\omega}}$. This simplified notation is used here instead of $\mathcal{H}(e^{i\omega})$.

$r$: Reference, a loop input, the desired plant output.

$e$: Controller input, the error signal, $e = r - y$.

$u$: Controller output, the command signal, a plant input.

$d$: Disturbance, a loop input, an unknown and unwanted plant input.

$x$: Plant input, $x = u + d$.

$y$: Plant output.

$\mathcal{H}_F^{input \to output}(z)$: Discrete-time transfer function for the closed-loop feedback system that connects an *input* (i.e. $r$ or $d$) to an *output* (i.e. $e$, $r$, $x$, or $y$).

$\blacksquare_P$: Pertaining to the plant system, i.e. the process to be controlled.

$\blacksquare_C$: Pertaining to the controller system, i.e. the digital system to be designed.

$\blacksquare_{\hat{C}}$: Pertaining to the controller system with the pure delay factored out.

$\blacksquare_F$: Pertaining to the (closed-loop) feedback system, containing the plant and the controller.

$K_\blacksquare$: Order of a linear time-invariant system, i.e. the number of poles.

$k$: Index into an array of coefficients, poles or components, $k = 0 \ldots K-1$.

$x(t)$: Input of a continuous-time system, e.g. the plant.

$y(t)$: Output of a continuous-time system, e.g. the plant.

$\mathcal{X}(s)$: Laplace transform of $x(t)$.

$\mathcal{Y}(s)$: Laplace transform of $y(t)$.

$\boldsymbol{A}$, $\boldsymbol{B}$, $\boldsymbol{C}$ & $\boldsymbol{D}$: State-space representation of a linear (time-invariant) system in continuous time.

$\boldsymbol{w}(t)$: Internal state vector of a linear state-space system in continuous time.

$\boldsymbol{w}(0)$: Initial state (at $t = 0$).

$\dot{\boldsymbol{w}}(t)$: Derivative (with respect to time) of the internal state vector of a linear state-space system in continuous time.

https://arxiv.org/abs/2211.09932

$\mathcal{W}(s)$: Laplace transform of $w(t)$.

$G(t)$: Fundamental matrix.

$\Phi(s)$: Laplace transform of $G(t)$.

$I$: Identity matrix, i.e. a zero matrix with ones along the main diagonal.

$H(t)$: Response matrix for a pulsed input that is held constant over one sampling period, a column vector for a single input system.

$x[n]$: Input of a generic discrete-time system, e.g. the discretized plant model.

$y[n]$: Output of a generic discrete-time system e.g. the discretized plant model.

$G$, $H$, $C$ & $D$: State-space representation of a linear (time-invariant) system $\mathcal{H}(z)$ in discrete time.

$w[n]$: Internal state vector of a discrete-time linear state-space system.

$\mathcal{W}(z)$, $\mathcal{X}(z)$ & $\mathcal{Y}(z)$: $\mathcal{Z}$-transforms of $w[n]$, $x[n]$ & $y[n]$.

$\bar{a}[k]$: The $k$th coefficient of modified denominator polynomial.

$p$: Desired position of (repeated) closed-loop poles on the real axis.

$\nu_F$, $\chi_P$ & $\mu_C$: Used to solve Diophantine equations for the controller coefficients.

$L(z)$: Loop function, the product of all transfer functions around the feedback loop, e.g. the plant and the controller.

$S(z)$: Sensitivity function, $S(z) = 1/\{1 + L(z)\}$.

$T(z)$: Complementary sensitivity function $T(z) = 1 - S(z)$.

$L(\omega)$: Frequency response of loop function, $L(\omega) = L(z)|_{z=e^{i\omega}}$.

$S(\omega)$: Frequency response of sensitivity function, $S(\omega) = S(z)|_{z=e^{i\omega}}$.

$T(\omega)$: Frequency response of complementary sensitivity function, $T(\omega) = T(z)|_{z=e^{i\omega}}$.

$\tilde{\varphi}$: Phase margin (in radians).

$\tilde{\omega}$: Gain cross-over frequency (in radians per sample) or the desired bandwidth of the loop.

$\tilde{f}$: Gain cross-over frequency (in cycles per sample) or the desired bandwidth of the loop.

$K_\psi$: Number of linear components used to form the controller via the frequency method.

$\psi_k(z)$: Discrete-time transfer function of the $k$th controller component.

$\psi_k(\omega)$: Frequency response of $\psi_k(z)$, $\psi_k(\omega) = \psi_k(z)|_{z=e^{i\omega}}$.

$c_k$: Linear coefficient that multiplies the $k$th controller component.

$L(\tilde{\omega})$: Desired (complex) value of $L(\omega)$ at $\tilde{\omega}$, $L(\tilde{\omega}) = e^{i(\tilde{\varphi}-\pi)}$.

$l$, $L$ & $c$: Used to solve for the controller coefficients via the frequency method.

$r_0$: The maximum radius (i.e. magnitude) of all closed-loop poles, for the nominal plant.

$r_1$: The maximum radius (i.e. magnitude) of all closed-loop poles, for the perturbed plant.

FIR: Finite Impulse Response.

IIR: Infinite Impulse Response.

## Introduction

This tutorial covers the essential principles of digital controller design. The reader is assumed to have been introduced to the Laplace transform, e.g. in a second-year mathematics course, and be comfortable with the $\mathcal{Z}$-transform, i.e. its discrete-time equivalent for sampled systems. Some exposure to feedback systems, as acquired in an engineering course on process control, digital signal processing, or analogue amplifiers, for instance, is also assumed. This document should be read in conjunction with books such as [1]&[2] for the theory of linear transforms and the modelling of linear systems, in continuous and discrete time; along with [3],[4]&[5] for more on digital controls; and [5]&[6] for robust control.

Unfortunately, many students are initially discouraged from acquiring a working knowledge of control theory by the seemingly abstract mathematics, e.g. complex functions of a complex argument in the s- and z-planes. Indeed, the meaning of the complex plane is obscure; however, for feedback systems it is the nexus that brings together the transient and steady-state responses in the time domain, the magnitude and phase response (at steady state) in the frequency domain, and ultimately the system realization using an analogue circuit or a digital computer. Thus, the abstract nature of the analysis is quickly forgotten when its practical value as an engineering tool is properly appreciated.

Feedback systems defy intuition, and complex analysis is essential to understand and predict their behaviour. Taking the output of a system and naively feeding it back into its input usually leads to unbounded growth (i.e. instability) and possibly disaster; however, when a digital controller is also placed in the loop, built using only a handful of delay, multiply and add operations, stability is maintained and behaviour emerges that mimics primitive intelligence, as the feedback system deftly tracks references or suppresses disturbances.

The essential theory required to derive the coefficients of a digital controller is presented here using a single worked example of potential practical utility – a tracking camera mount. Two complementary design procedures are discussed – the *polynomial* method and the *frequency* method. They are used to highlight different aspects of control theory, or alternative ways of looking at the control problem. Both methods exploit the same hypothetical plant model (i.e. the process to be controlled); however, the former method focusses on system poles and zeros in the complex z-plane; whereas the latter method uses stability margins and bandwidth, along the frequency axis (i.e. around the unit circle in the complex z-plane).

## System overview

An object tracking system is used here as a worked example to illustrate two controller design methods and to elucidate the mathematics of linear time-invariant signals-and-systems. The hypothetical system seeks to track a designated target in the azimuth coordinate and keep it near the centre of the field of view of a video camera that is attached to a motorized camera mount (see Figure 1). The target and camera are assumed to be co-planar, thus elevation is not considered here. The frame rate ($F_s$) of the camera is 100 Hz. Its zoom, focus and aperture settings are assumed to be appropriately set and held constant. An image processor (e.g. a blob detector) is used to determine the azimuth difference between the target and the camera in every image frame (the details of this algorithm are not considered here). The primary design objective is to derive the coefficients of a digital feedback controller that automatically issues appropriate commands to the mount so that the target is continuously observed by the video camera. The system should also have a video stabilization capability, for stationary objects that remain in the field of view. The image processor and mount controller are hosted on a personal computer. Control commands output from the computer (at a rate of 100 Hz) are sent to the mount via a serial cable and video frames from the camera are input to the computer via an ethernet cable. The digital controller operates at the frame rate of the video stream and the sampling times are assumed to be approximately synchronized.

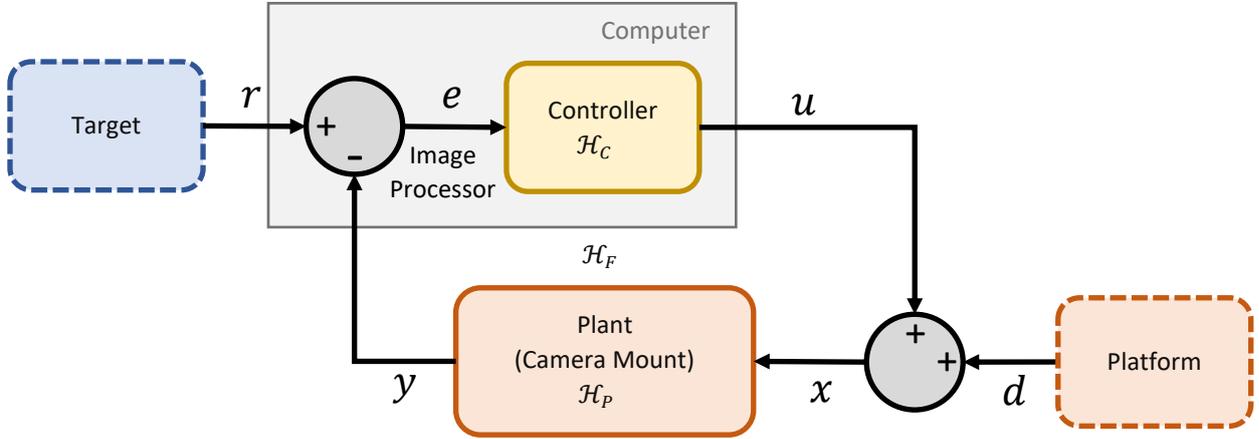

*Figure 1. System block diagram showing the target (an object to be tracked by the feedback system), the controller (a low-order digital filter to be designed), the plant (a camera mount), and the Platform (a vehicle or structure on which the camera mount is attached).*

$r$: Reference, a loop input, the desired plant output (target azimuth).

$e$: Controller input, the error signal (the difference between the target azimuth and the camera azimuth, as determined by an object detector or image processor operating on a given image frame, that is collected by the video camera).

$u$: Controller output, the command signal (azimuth rate), an input to the plant.

$d$: Disturbance, a loop input, an unwanted plant input (an unknown perturbation of the camera in azimuth, caused by a rotation of the sensing platform, for example).

$x$: Plant input (azimuth rate).

$y$: Plant output (camera azimuth).

$\mathcal{H}_C$: Transfer function of <u>C</u>ontroller system.

$\mathcal{H}_P$: Transfer function of <u>P</u>lant system.

$\mathcal{H}_F$: Transfer function of <u>F</u>eedback system.

In Figure 1, the error signal ($e$) is the difference between reference ($r$) and the plant output ($y$), i.e. $e = r - y$. Additional errors introduced by image sensing and processing are not considered here, caused by, changing target aspect, clutter in the scene, reflection/refraction/obscuration of light, pointing errors, for example. This simplifies the system model substantially because additional transfer functions (and error inputs) are not required at the plant output and the reference input. For instance, the "sensor" transfer function and the "sensor noise" input considered in [6], are not explicitly modelled here.

## Plant model

The Fourier transform $\mathcal{F}\{\blacksquare\}$ of a system's impulse response describes how the system responds to sinusoidal inputs. The sinusoids are of infinite extent and the system response, which is expressed simply as a magnitude scaling and a phase shift at all measurable frequencies, is for the system at steady state, i.e. after infinite time has elapsed. Frequency-domain representations are an invaluable engineering tool; however in control-system engineering, the transient behaviour of the system, as it responds to sudden input changes, then *approaches* steady state, is also of great interest. The complex s-plane reached via the Laplace transform $\mathcal{L}\{\blacksquare\}$ has two axes: a vertical imaginary axis that corresponds to frequency (i.e. phase rate, in radians per second) and a horizontal real axis that corresponds to rate (in reciprocal seconds). The imaginary axis contains the same steady-state information as the frequency axis of the Fourier transform; however, the extra degree of freedom provided by the real axis captures the rate of (exponential) growth or decay of a system in the transient regime.

The Laplace transform $\mathcal{L}\{\blacksquare\}$ maps a function on a one-dimensional time axis, into a two-dimensional complex s-plane with real and imaginary axes; for example, $\mathcal{L}\{h(t)\} = \mathcal{H}(s)$ where $h(t)$ is a system's impulse response and $\mathcal{H}(s)$ is a system's transfer-function [1]&[2]. Coordinates in the complex s-plane, reached via the Laplace transform, are represented using $s = \sigma + i\Omega$, where $i = \sqrt{-1}$ is the imaginary unit, $\Omega$ is the imaginary coordinate (i.e. along the phase-rate or frequency axis of the s-plane, in units of radians per second) and $\sigma$ is the real coordinate (i.e. along the rate axis of the s-plane, in units of reciprocal seconds). The transfer function of a linear time-invariant system is a rational function in $s$; i.e. it is a ratio of polynomials $\mathcal{H}(s) = \mathcal{B}(s)/\mathcal{A}(s)$ and the roots of the denominator and numerator polynomials, i.e. the complex values of $s$ for which $|\mathcal{A}(s)| = 0$ and $|\mathcal{B}(s)| = 0$, are referred to as poles and zeros, respectively. Note that $\mathcal{H}(s)$ is a complex function over a complex plane; furthermore, $|\mathcal{H}(s)| = \infty$ and $|\mathcal{H}(s)| = 0$ at the locations of poles and zeros, respectively, where $|\blacksquare|$ is the magnitude operator.

The physical significance of pole (and zero) coordinates in the complex s-plane is best illustrated by example. Consider the complex exponential $h(t) = e^{\sigma t + i\Omega t}$. It has an exponential envelope that modulates a sinusoidal carrier, i.e. $h(t) = e^{\sigma t} e^{i\Omega t}$ and it has a Laplace transform of $\mathcal{L}\{h(t)\} = \mathcal{H}(s) = 1/(s - \sigma - i\Omega)$ with a pole at $s = \sigma + i\Omega$ where $|\mathcal{H}(s)| = \infty$. The two degrees of freedom for the impulse response $h(t)$ are represented by the $\sigma$ (rate) and $\Omega$ (phase-rate or frequency) axes respectively: $\sigma$ determines the rate of exponential *growth* (for $\sigma > 0$) or *decay* (for $\sigma < 0$) and $\Omega$ determines the oscillation frequency.

On earth and in its neighbourhood, stable systems (with poles on the left-hand side of the s-plane, where $\sigma < 0$) prevail; however, unstable systems (with poles on the right-hand side of the s-plane, where $\sigma > 0$) are of particular interest to humans. Either way, all timescales (i.e. $1/|\sigma|$) are encountered, from the infinitesimal (i.e. $|\sigma| \to \infty$) to the infinite (i.e. $|\sigma| \to 0$). It is also worth noting that fast dynamics (with a rate of $\sigma$ and a phase rate of $\Omega$) are difficult to measure, model, and manipulate, and that low-order low-frequency models with poles near the origin of the s-plane (i.e. $|\sigma + i\Omega| \approx 0$) are often sufficient and usually recommended [5].

Roots of $\mathcal{H}(s)$ that are close to the imaginary axis (i.e. where $|\sigma| \approx 0$) are said to be "dominant" because they broadly determine (or describe) process dynamics and form reasonable low-order models. They have a pronounced impact on the frequency response of the system $\mathcal{H}(\Omega)$ which is determined by evaluating $\mathcal{H}(s)$ along the imaginary axis, i.e. $\mathcal{H}(\Omega) = \mathcal{H}(s)|_{s=i\Omega}$; thus its impulse response $h(t)$, which may also be determined by taking the inverse Fourier transform of $\mathcal{H}(\Omega)$, i.e. $h(t) = \mathcal{F}^{-1}\{\mathcal{H}(\Omega)\}$, if defined. Approximate plant models formed from dominant poles (and zeros) also simplify controller design, analysis, implementation, integration, deployment and maintenance.

A simple second-order plant model is assumed for the camera-mount plant considered here, for analytical simplicity and robust stability. The output of the plant is the azimuth of the camera in

https://arxiv.org/abs/2211.09932degrees. The camera is assumed to rotate freely and without limit. The camera mount accepts azimuth-rate commands (degrees per second) with an unknown/unspecified maximum rotation rate.

A simple integrator that has one pole at the origin of the complex s-plane (i.e. at $s = 0$) is a reasonable first-order approximation of the plant. For $t \geq 0$, an integrator has an impulse response of $h(t) = 1$ and a (unit) step response of $h(t) = t$. However, the plant response also exhibits a brief lag, i.e. some 'slack' that 'smooths' sudden input changes. A second-order plant model with one pole at the origin and another on the negative real axis is therefore a better approximation.

The transfer function of a first-order system with one pole on the real axis is $\mathcal{H}(s) = 1/(s - \sigma)$ and its impulse response is $h(t) = e^{\sigma t}$ (for $t \geq 0$). For a first-order *lag* with an exponential *decay* (i.e. $\sigma < 0$) the mean impulse-response duration is $-1/\sigma$ in units of seconds.

The continuous-time second-order transfer-function of the plant $\mathcal{H}_P(s)$ is the product of a first-order integrator and a first-order lag. It is a ratio of complex polynomials in *s* i.e.

$\mathcal{H}_P(s) = \mathcal{B}_P(s)/\mathcal{A}_P(s)$ with (1a)

$A_P(s) = s(s - \sigma)$ . (1b)

The transfer function is normalised so that the rate of change of the output at steady state (i.e. as $t \to \infty$) for a unit step input is equal to unity. Let $\mathcal{X}(s)$ and $\mathcal{Y}(s)$ be the Laplace transforms of the input and output, respectively, with $\mathcal{X}(s) = 1/s$ for a unit step input. By definition, we have $\mathcal{Y}(s) = \mathcal{H}_P(s)\mathcal{X}(s)$ and for the derivative of the output we have $s\mathcal{Y}(s) = s\mathcal{H}_P(s).\mathcal{X}(s)$. The final-value theorem states that $\lim_{t \to \infty} h(t) = \lim_{s \to 0} s\mathcal{H}(s)$. Application of this theorem to the derivative of the output therefore yields

$\lim_{s \to 0} s^2 \mathcal{H}_P(s)\mathcal{X}(s) = 1$ . (2)

Substituting $\mathcal{H}_P(s) = \frac{B_P(s)}{s(s-\sigma)}$ and $\mathcal{X}(s) = 1/s$ into (2), i.e.

$\lim_{s \to 0} \frac{B_P(s)}{(s-\sigma)} = 1$ (3)

and evaluating the limit, yields $\mathcal{B}_P(s) = -\sigma$; thus

$\mathcal{H}_P(s) = \frac{\sigma}{s(\sigma-s)}$ . (4)

For the tracking camera mount in Figure 1, plant models with smaller rate parameters (i.e. $\sigma$) are appropriate for larger and/or heavier mounts.

Linear state-space models are a convenient representation of continuous-time and discrete-time systems alike. They are used here to compute the plant output for a sequence of contiguous input pulses (i.e. 'commands') that are output by the controller [3].

The dynamics of a linear system in continuous time are concisely represented using a differential equation of the form

$\dot{\boldsymbol{w}}(t) = \boldsymbol{A}\boldsymbol{w}(t) + \boldsymbol{B}x(t)$ and (5a)

$y(t) = \boldsymbol{C}\boldsymbol{w}(t) + \boldsymbol{D}x(t)$ (5b)

(where $\dot{\blacksquare}$ is the first derivative operator, with respect to time)

or in the s-domain using

$$s\mathcal{W}(s) = A\mathcal{W}(s) + B\mathcal{X}(s) \text{ and} \tag{6a}$$

$$\mathcal{Y}(s) = C\mathcal{W}(s) + D\mathcal{X}(s). \tag{6b}$$

For our second-order plant we have the following differential equations that govern its dynamics:

$$A = \begin{bmatrix} 0 & 1 \\ 0 & \sigma \end{bmatrix}, B = \begin{bmatrix} 0 \\ 1 \end{bmatrix}, C = [-\sigma \quad 0], D = 0 \text{ where} \tag{7}$$

$w$ is a $2 \times 1$ state vector with azimuth and azimuth-rate elements (from top to bottom).

The time evolution of a linear state-space system ($A$, $B$, $C$, and $D$), that has an initial state of $w(0)$ and is driven by an input of $x(t)$, is found by solving (5) via (6) to obtain the $G(t)$ and $H(t)$ operators.

*In the absence of inputs*, and from an initial state of $w(0)$, the solution has the following form:

$$w(t) = G(t)w(0) \text{ where} \tag{8a}$$

$G(t)$ is a $2 \times 2$ (so-called 'fundamental') matrix with elements, (8b)
that are a function of time, determined using

$$G(t) = \mathcal{L}^{-1}\{\Phi(s)\} \text{ where} \tag{8c}$$

$$\Phi(s) = (sI - A)^{-1} \text{ with} \tag{8d}$$

$I$ being the $2 \times 2$ identity matrix.

For our second-order plant model in (7) we have the solution

$$G(t) = \begin{bmatrix} 1 & (e^{\sigma t} - 1)/\sigma \\ 0 & e^{\sigma t} \end{bmatrix}. \tag{9}$$

*From a zero initial state*, and for a unit step input, i.e. $x(t) = 1$ for $t \geq 0$ or $\mathcal{X}(s) = s^{-1}$, the solution has the following form:

$$w(t) = H(t)x(t) \text{ where} \tag{10a}$$

$H(t)$ is a $2 \times 1$ input response vector (10b)
with elements, that are a function of time, determined using

$$H(t) = \mathcal{L}^{-1}\{\Phi(s)Bs^{-1}\}. \tag{10c}$$

For our second-order plant model defined using (7) we have the solution

$$H(t) = \begin{bmatrix} e^{\sigma t}/\sigma^2 - t/\sigma - 1/\sigma^2 \\ e^{\sigma t}/\sigma - 1/\sigma \end{bmatrix}. \tag{11}$$

For a linear time-invariant system with a non-zero initial state and a non-zero input the solutions are simply added, yielding

$$w(t) = G(t)w(0) + H(t)x(t). \tag{12}$$

Alternatively, for an input that is held constant over one sampling period of $T_s = 1/F_s$ seconds, from an arbitrary time of $t$

$$w(t + T_s) = Gw(t) + Hx(t) \text{ where} \tag{13a}$$

$$G = G(t)|_{t=T_s} \text{ and} \tag{13b}$$

$$H = H(t)|_{t=T_s} . \tag{13c}$$

Let $t = T_s n$, where $n$ is an integer sample index (for $n = 0 \ldots \infty$). The discrete-time state equations are now simply

$$w[n + 1] = Gw[n] + Hx[n] \tag{14a}$$

and the output of the plant *at the sample times* is determined using

$$y[n] = Cw[n] + Dx[n] . \tag{14b}$$

Note that at the $\sigma \to 0$ limit we have two poles at $s = 0$, i.e. a double integrator, e.g. for a Newtonian system with an acceleration input and a position output, yielding the more familiar discrete-time state equations (after appropriate re-normalization)

$$G = \begin{bmatrix} 1 & T_s \\ 0 & 1 \end{bmatrix} \tag{15a}$$

$$H = \begin{bmatrix} \frac{1}{2}T_s^2 \\ T_s \end{bmatrix} \text{ and} \tag{15c}$$

$$C = \begin{bmatrix} 1 & 0 \end{bmatrix} \text{ with the} \tag{15d}$$

$k = 0$ & $k = 1$ elements of the $w[n]$ state vector corresponding to angular position and angular velocity.

In open-loop and closed-loop configurations alike, the output of the controller is input to the plant and its response to a sequence of control commands that are held constant over one sampling interval (i.e. using a so-called "zero-order hold") is simply determined using the linear state-space recursion in (14).

The discrete-time transfer function $\mathcal{H}_P(z)$ for the sampled plant, that is connected to the digital controller via a zero-order hold, is reached by taking the $\mathcal{Z}$-transform of (14) yielding

$$z\mathcal{W}(z) = G\mathcal{W}(z) + H\mathcal{X}(z) \text{ and} \tag{16a}$$

$$\mathcal{Y}(z) = C\mathcal{W}(z) + D\mathcal{X}(z) . \tag{16b}$$

The $\mathcal{Z}$-transform of the state vector $\mathcal{W}(z)$ is eliminated by rearranging (16a)

$$(zI - G)\mathcal{W}(z) = H\mathcal{X}(z) \text{ then} \tag{17a}$$

$$\mathcal{W}(z) = (zI - G)^{-1}H\mathcal{X}(z) \text{ which is then substituted into (16b) yielding} \tag{17b}$$

$$\mathcal{Y}(z) = C(zI - G)^{-1}H\mathcal{X}(z) + D\mathcal{X}(z) \text{ or} \tag{17c}$$

$$\mathcal{Y}(z) = \{C(zI - G)^{-1}H + D\}\mathcal{X}(z) . \tag{17d}$$

As the discrete-time transfer function of a system is defined using $\mathcal{H}(z) = \mathcal{Y}(z)/\mathcal{X}(z)$ or $\mathcal{Y}(z) = \mathcal{H}(z)\mathcal{X}(z)$, from (17d) we have

$$\mathcal{H}_P(z) = C\Phi(z)H + D \text{ where} \tag{18a}$$

$$\Phi(z) = (zI - G)^{-1}. \tag{18b}$$

Note that the z-domain derivation of $\Phi(z)$ in (17) & (18) is analogous to the procedure used to derive $\Phi(s)$ in the s-domain in (8).

Substitution of **G**, **H**, **C**, and **D** for our second-order plant model into (18) yields

$$\mathcal{H}_P(z) = \mathcal{B}_P(z)/\mathcal{A}_P(z) \text{ with} \tag{19a}$$

$$\mathcal{B}_P(z) = \frac{1}{\sigma}(\sigma T_s - e^{\sigma T_s} + 1)z - \frac{1}{\sigma}(\sigma T_s e^{\sigma T_s} - e^{\sigma T_s} + 1) \text{ and} \tag{19b}$$

$$\mathcal{A}_P(z) = z^2 - (1 + e^{\sigma T_s})z + e^{\sigma T_s} . \tag{19c}$$

For this discrete-time transfer function the order of the numerator polynomial $\mathcal{B}_P(z)$ is less than the order of the denominator polynomial $\mathcal{A}_P(z)$, thus **D** = 0. Such transfer functions, e.g. for Newtonian mechanical systems are classified as *strictly proper*. Systems that produce instantaneous outputs from inputs, e.g. electrochemical systems in nature or digital filters & controls in computers, may have equal numerator and denominator orders, thus **D** ≠ 0. Such transfer functions are classified as *proper*.

Feedback drives the tracking error (i.e. the error signal, $e$) relative to a desired reference point ($r$, e.g. the target direction) towards zero. It also drives the net plant input ($x$) towards zero for unwanted & unknown disturbances ($d$, e.g. a platform rotation). Without plant uncertainty, feedback is unnecessary and open-loop configurations would be sufficient for the target tracking function of the feedback system in (Figure 1); and without disturbances, the image stabilization function would be unnecessary. Two complementary (*polynomial* and *frequency*) methods for the design of digital controller coefficients are described in the following section. The controller generates an appropriate command signal ($u$) from an observed error signal ($e$) that is generated inside a feedback loop.

## Controller design

Feedback structures allow partially known processes to be adequately controlled using only approximate models (e.g. second-order). Many natural and mechanical processes have one or two dominant poles near the origin of the complex s-plane for a low-pass frequency response, and for such systems, Proportional Integral Derivative (PID) controls are ideal. A digital PID controller has a pole at $z = 0$ (a one-sample delay) and a pole at $z = 1$ due to the integrator, which applies a very high gain at very low frequencies (infinite at dc). The frequency method uses the PID structure as a 'compensator' to shape the response of the loop: the 'P' term applies uniform gain over all frequencies, the 'D' term (with a zero at $z = 1$) amplifies medium-to-high frequencies, and the 'I' term applies a very high gain at very low frequencies. The digital PI and PID structures are first- and second-order filters (respectively) with an infinite impulse response (IIR) due to the integrator; whereas the PD structure is a simple first-order filter with a finite impulse response (FIR). An integrator is also incorporated into the polynomial method as a special case to improve the steady-state response. For unstable or oscillatory plants of low order, more general second-order IIR structures with complex poles are usually required. The polynomial and frequency methods both accommodate such cases; however, only PD and PID structures are covered here for the frequency method as they are sufficient for the assumed (stable and non-oscillatory) plant model.

### Polynomial design

The polynomial method is a form of controller design by pole placement, i.e. the digital controller is designed to precisely position the poles of the feedback system in which it is placed [3],[4]&[5]. The poles of a transfer function, e.g. $\mathcal{H}(z)$, determine the rate of decay or growth and the rate of phase change (i.e. frequency) of its impulse response, e.g. $h[m]$ for $m = 0 \ldots \infty$. The output $y[n]$ or response in time for any input (e.g. pulse, step, or ramp etc) is then determined by convolving the input $x[n]$ with $h[m]$ using $y[n] = \sum_{m=0}^{\infty} h[m] \, x[n-m]$.

The poles of a system determine its resonant modes and the zeros scale the magnitude and shift the phase of oscillation for those modes. Thus, the poles of a system are the primary representation of its behaviour, and its zeros are of secondary importance. All degrees of design freedom are therefore used to place the poles of the closed-loop system. Furthermore, the transfer functions derived below for the feedback system in Figure 1 show that the poles of the feedback system are common to all input-output relationships around the loop and that only their zeros differ. The polynomial method places poles for more-or-less the desired response. In principle, carefully placed zeros may refine the response for a particular input-output relationship of interest, e.g. $r \to y$ for a reference-tracking servomechanism or $d \to y$ for a disturbance-rejecting regulator. However, there is no point finessing theoretical transient responses if the behaviour of the plant is not known precisely and if robust stability is a priority [5].

For a given plant with $\mathcal{H}_P(z) = \mathcal{B}_P(z)/\mathcal{A}_P(z)$ the polynomial method determines the coefficients for the digital controller $\mathcal{H}_C(z) = \mathcal{B}_C(z)/\mathcal{A}_C(z)$ so that the feedback system $\mathcal{H}_F(z) = \mathcal{B}_F(z)/\mathcal{A}_F(z)$ has its poles placed at the desired locations, i.e. for $\mathcal{A}_F(z)$ specified. All input/output relationships around a feedback loop have the same denominator polynomial $\mathcal{A}_F(z)$, i.e. they have common poles, although the numerator polynomials $\mathcal{B}_F(z)$ are not necessarily the same. The denominator and numerator polynomials describe the parts of the system's behaviour that are due to feedback and 'feedforward', respectively. The denominator for all input-output relationships around the loop is equal to unity plus the product of all transfer functions around the loop whereas the numerator is determined only by the transfer functions that are in between the input and the output points. For the feedback system in Figure 1, the closed-loop transfer functions that connect any system input, e.g. $r$ or $d$, to any system output, e.g. $e$, $u$, $x$, or $y$, are

$$\mathcal{H}_F^{r \to e}(z) = 1/\{1 + \mathcal{H}_C(z)\mathcal{H}_P(z)\} \tag{20a}$$

$$\mathcal{H}_F^{r \to u}(z) = \mathcal{H}_C(z)/\{1 + \mathcal{H}_C(z)\mathcal{H}_P(z)\} \tag{20b}$$

$$\mathcal{H}_F^{r \to y}(z) = \mathcal{H}_C(z)\mathcal{H}_P(z)/\{1 + \mathcal{H}_C(z)\mathcal{H}_P(z)\} \tag{20c}$$

$$\mathcal{H}_F^{d \to x}(z) = 1/\{1 + \mathcal{H}_C(z)\mathcal{H}_P(z)\} \tag{20d}$$

$$\mathcal{H}_F^{d \to y}(z) = \mathcal{H}_P(z)/\{1 + \mathcal{H}_C(z)\mathcal{H}_P(z)\} \text{ and} \tag{20e}$$

$$\mathcal{H}_F^{d \to u}(z) = -\mathcal{H}_C(z)\mathcal{H}_P(z)/\{1 + \mathcal{H}_C(z)\mathcal{H}_P(z)\} \,. \tag{20f}$$

For all $\mathcal{H}_F(z)$ in (20), the denominator polynomial has the form $\mathcal{A}_F(z) = 1 + L(z)$, where $L(z)$ is the *loop function*, which is the product of all transfer functions around/inside the feedback loop. For the system in Figure 1 the loop function is

$$L(z) = \mathcal{H}_C(z)\mathcal{H}_P(z) \,. \tag{21}$$

For more general systems $L(z)$ may also include other transfer functions e.g. for actuators, sensors, antennas, transmitters or receivers. The loop function $L(z)$ has a simpler form than the transfer function of the feedback loop $\mathcal{H}_F(z)$ which makes it very useful for design and analysis purposes. It also determines the properties of the feedback loop that is formed around it.

Substitution of $\mathcal{H}_C(z) = \mathcal{B}_C(z)/\mathcal{A}_C(z)$ and $\mathcal{H}_P(z) = \mathcal{B}_P(z)/\mathcal{A}_P(z)$ into (20) above and simplifying yields

$$\mathcal{H}_F^{r \to e}(z) = \mathcal{A}_C(z)\mathcal{A}_P(z)/\{\mathcal{A}_C(z)\mathcal{A}_P(z) + \mathcal{B}_C(z)\mathcal{B}_P(z)\} \tag{22a}$$

$$\mathcal{H}_F^{r \to u}(z) = \mathcal{B}_C(z)\mathcal{A}_P(z)/\{\mathcal{A}_C(z)\mathcal{A}_P(z) + \mathcal{B}_C(z)\mathcal{B}_P(z)\} \tag{22b}$$

$$\mathcal{H}_F^{r \to y}(z) = \mathcal{B}_C(z)\mathcal{B}_P(z)/\{\mathcal{A}_C(z)\mathcal{A}_P(z) + \mathcal{B}_C(z)\mathcal{B}_P(z)\} \tag{22c}$$

$$\mathcal{H}_F^{d \to x}(z) = \mathcal{A}_C(z)\mathcal{A}_P(z)/\{\mathcal{A}_C(z)\mathcal{A}_P(z) + \mathcal{B}_C(z)\mathcal{B}_P(z)\} \tag{22d}$$

$$\mathcal{H}_F^{d \to y}(z) = \mathcal{A}_C(z)\mathcal{B}_P(z)/\{\mathcal{A}_C(z)\mathcal{A}_P(z) + \mathcal{B}_C(z)\mathcal{B}_P(z)\} \text{ and} \tag{22e}$$

$$\mathcal{H}_F^{d \to u}(z) = -\mathcal{B}_C(z)\mathcal{B}_P(z)/\{\mathcal{A}_C(z)\mathcal{A}_P(z) + \mathcal{B}_C(z)\mathcal{B}_P(z)\}. \tag{22f}$$

Note that $\mathcal{H}_F^{r \to x}(z) = \mathcal{H}_F^{r \to u}(z)$ and $\mathcal{H}_F^{d \to e}(z) = -\mathcal{H}_F^{d \to y}(z)$. Note also that when both references and disturbances are applied, the response of the feedback system for a given output of interest (i.e. $e$, $u$, $x$ or $y$) is simply determined by adding the outputs due to both inputs (i.e. $r$ & $d$).

Within the feedback loop, the numerator and denominator polynomials of the controller and the plant are 'intermingled' to form poles and zeros of the various input-output relationships around the loop. The coefficients of the digital controller, i.e. the $\mathcal{B}_C(z)$ and $\mathcal{A}_C(z)$ polynomials of $\mathcal{H}_C(z)$, are chosen so that the poles of $\mathcal{H}_F(z)$, i.e. the roots of the $\mathcal{A}_C(z)\mathcal{A}_P(z) + \mathcal{B}_C(z)\mathcal{B}_P(z)$ polynomial in (22), are equal to the poles of the desired closed-loop transfer function, i.e.

$$\mathcal{A}_C(z)\mathcal{A}_P(z) + \mathcal{B}_C(z)\mathcal{B}_P(z) = \mathcal{A}_F(z). \tag{23}$$

General polynomial problems of this type are referred to as Diophantine equations [3],[4]&[5].

In principle, the poles of $\mathcal{H}_F(z)$ maybe chosen arbitrarily and positioned anywhere inside the unit circle in the complex z-plane. If complex poles are used, they should appear in conjugate pairs. Real poles on the (0,1) interval provide well-damped (non-oscillatory) responses. Using repeated real poles, all placed at $p$, simplifies the design process and is adequate for the plant model considered here. As $p \to 0$, the bandwidth of the closed-loop response increases, for a faster (i.e. 'compressed') transient response. For a tracking camera mount, a wider bandwidth is required for manoeuvring targets (at the expense of greater white-noise amplification). A pure delay results when $p = 0$, for a so-called "deadbeat" response when the plant model is perfect [3]; however, aiming for this ideal response in theory, usually requires aggressive control action and it may result in an unstable feedback loop when the plant model is imperfect, which is almost always the case. As $p \to 1$, the bandwidth of the closed-loop response decreases, for a slower (i.e. 'sluggish') transient response.

For the lagged integrator plant in (19) we have

$$\mathcal{B}_P(z) = b_P[0]z^2 + b_P[1]z + b_P[2] \text{ where} \tag{24a}$$

$$b_P[0] = 0, \ b_P[1] = \frac{1}{\sigma}(\sigma T_s - e^{\sigma T_s} + 1) \text{ and } b_P[2] = -\frac{1}{\sigma}(\sigma T_s e^{\sigma T_s} - e^{\sigma T_s} + 1) \text{ with} \tag{24b}$$

$$\mathcal{A}_P(z) = a_P[0]z^2 + a_P[1]z + a_P[2] \text{ where} \tag{24c}$$

$$a_P[0] = 1, \ a_P[1] = -(1 + e^{\sigma T_s}) \text{ and } a_P[2] = e^{\sigma T_s}. \tag{24d}$$

### Third-order polynomial design

Placing the plant inside a feedback loop with a controller allows the closed-loop poles to be placed at $p$. For a second-order plant model ($K_P = 2$), with a third-order controller, (23) indicates that we have a fifth-order closed-loop system ($K_F = K_P + K_C = 5$) with

$$\mathcal{A}_F(z) = (z - p)^5 = a_F[0]z^5 + a_F[1]z^4 + a_F[2]z^3 + a_F[3]z^2 + a_F[4]z + a_F[5] \tag{25a}$$

where

$$a_F[0] = 1, \ a_F[1] = -5p, \ a_F[2] = 10p^2, \ a_F[3] = -10p^3, \ a_F[4] = 5p^4 \text{ and } a_F[5] = -p^5. \tag{25b}$$

The controller contains a delay of one sample to support the computer implementation of the loop in Figure 1, i.e. $\mathcal{H}_C(z)$ contains a factor of $1/z$ to delay the error signal $e$. The controller, with the required delay, has an adjustable gain (of $\gamma_C$), two adjustable zeros (at $\beta_0$ & $\beta_1$) and two adjustable poles (at $\alpha_0$ & $\alpha_1$). It therefore has the following form:

$$\mathcal{B}_C(z) = \gamma_C(z-\beta_0)(z-\beta_1) = b_C[0]z^3 + b_C[1]z^2 + b_C[2]z + b_C[3] \text{ with} \tag{26a}$$

$$b_C[0] = 0 \text{ and} \tag{26b}$$

$$\mathcal{A}_C(z) = z(z-\alpha_0)(z-\alpha_1) = a_C[0]z^3 + a_C[1]z^2 + a_C[2]z + a_C[3] \text{ with} \tag{26c}$$

$$a_C[0] = 1 \text{ and } a_C[3] = 0. \tag{26d}$$

We therefore have (five) unknown controller coefficients $a_C$ & $b_C$ in (26) that are used to combine the plant coefficients $a_P$ & $b_P$ in (24), according to (23), so that the (five) denominator coefficients $a_F$, for the desired closed-loop response in (25), are obtained. For the sum of polynomial products in (23), the following set of linear equations result:

$$
\begin{aligned}
a_F[0]z^5 &= a_P[0]z^2 a_C[0]z^3 \\
a_F[1]z^4 &= a_P[1]z^1 a_C[0]z^3 + a_P[0]z^2 a_C[1]z^2 && + b_P[0]z^2 b_C[1]z^2 \\
a_F[2]z^3 &= a_P[2]z^0 a_C[0]z^3 + a_P[1]z^1 a_C[1]z^2 + a_P[0]z^2 a_C[2]z^1 + b_P[1]z^1 b_C[1]z^2 + b_P[0]z^2 b_C[2]z^1 \\
a_F[3]z^2 &= \phantom{a_P[2]z^0 a_C[0]z^3 +} a_P[2]z^0 a_C[1]z^2 + a_P[1]z^1 a_C[2]z^1 + b_P[2]z^0 b_C[1]z^2 + b_P[1]z^1 b_C[2]z^1 + b_P[0]z^2 b_C[3]z^0 \\
a_F[4]z^1 &= \phantom{a_P[2]z^0 a_C[0]z^3 + a_P[1]z^1 a_C[1]z^2 +} a_P[2]z^0 a_C[2]z^1 \phantom{+ b_P[2]z^0 b_C[1]z^2} + b_P[2]z^0 b_C[2]z^1 + b_P[1]z^1 b_C[3]z^0 \\
a_F[5]z^0 &= \phantom{a_P[2]z^0 a_C[0]z^3 + a_P[1]z^1 a_C[1]z^2 + a_P[0]z^2 a_C[2]z^1 + b_P[1]z^1 b_C[1]z^2 + b_P[0]z^2 b_C[2]z^1} b_P[2]z^0 b_C[3]z^0
\end{aligned}
$$
$$\tag{27a}$$

Substitute $b_P[0] = 0$, $a_P[0] = 1$, $a_F[0] = 1$ and combine factors of $z$:

$$
\begin{aligned}
z^5 &= a_C[0]z^5 \\
a_F[1]z^4 &= a_P[1]a_C[0]z^4 + a_C[1]z^4 \\
a_F[2]z^3 &= a_P[2]a_C[0]z^3 + a_P[1]a_C[1]z^3 + a_C[2]z^3 + b_P[1]b_C[1]z^3 \\
a_F[3]z^2 &= \phantom{a_P[2]a_C[0]z^3 +} a_P[2]a_C[1]z^2 + a_P[1]a_C[2]z^2 + b_P[2]b_C[1]z^2 + b_P[1]b_C[2]z^2 \\
a_F[4]z^1 &= \phantom{a_P[2]a_C[0]z^3 + a_P[1]a_C[1]z^3 +} a_P[2]a_C[2]z^1 \phantom{+ b_P[2]b_C[1]z^2} + b_P[2]b_C[2]z^1 + b_P[1]b_C[3]z^1 \\
a_F[5]z^0 &= \phantom{a_P[2]a_C[0]z^3 + a_P[1]a_C[1]z^3 + a_P[2]a_C[1]z^2 + b_P[1]b_C[1]z^2 + b_P[2]b_C[2]z^1} b_P[2]b_C[3]z^0
\end{aligned}
$$
$$\tag{27b}$$

Substitute $a_C[0] = 1$ so the first row can be eliminated and subtract the first 'column' from both sides:

$$
\begin{aligned}
a_F[1]z^4 - a_P[1]z^4 &= a_C[1]z^4 \\
a_F[2]z^3 - a_P[2]z^3 &= a_P[1]a_C[1]z^3 + a_C[2]z^3 + b_P[1]b_C[1]z^3 \\
a_F[3]z^2 &= a_P[2]a_C[1]z^2 + a_P[1]a_C[2]z^2 + b_P[2]b_C[1]z^2 + b_P[1]b_C[2]z^2 \\
a_F[4]z^1 &= \phantom{a_P[2]a_C[1]z^2 +} a_P[2]a_C[2]z^1 \phantom{+ b_P[2]b_C[1]z^2} + b_P[2]b_C[2]z^1 + b_P[1]b_C[3]z^1 \\
a_F[5]z^0 &= \phantom{a_P[2]a_C[1]z^2 + a_P[1]a_C[2]z^2 + b_P[2]b_C[1]z^2 + b_P[1]b_C[2]z^2} b_P[2]b_C[3]z^0
\end{aligned}
$$
$$\tag{27c}$$

Or using matrix/vector notation,

$$\boldsymbol{v}_F = \boldsymbol{\chi}_P \boldsymbol{\mu}_C \text{ with} \tag{28a}$$

$$\boldsymbol{v}_F = \begin{bmatrix} a_F[1]-a_P[1] \\ a_F[2]-a_P[2] \\ a_F[3] \\ a_F[4] \\ a_F[5] \end{bmatrix}, \boldsymbol{\chi}_P = \begin{bmatrix} 1 & 0 & 0 & 0 & 0 \\ a_P[1] & 1 & b_P[1] & 0 & 0 \\ a_P[2] & a_P[1] & b_P[2] & b_P[1] & 0 \\ 0 & a_P[2] & 0 & b_P[2] & b_P[1] \\ 0 & 0 & 0 & 0 & b_P[2] \end{bmatrix}, \boldsymbol{\mu}_C = \begin{bmatrix} a_C[1] \\ a_C[2] \\ b_C[1] \\ b_C[2] \\ b_C[3] \end{bmatrix} \tag{28b}$$

which is readily solved for $\boldsymbol{\mu}_C$ using

$$\boldsymbol{\mu}_C = \boldsymbol{\chi}_P^{-1} \boldsymbol{v}_F \text{ where} \tag{28c}$$

$\blacksquare^{-1}$ is a matrix inverse operation. (28d)

The poles of the controller are equal to the roots of $\mathcal{A}_C(z)$ and their positions should be carefully checked before the solution is accepted. They may be located anywhere in the z-plane (as a conjugate pair if complex) and there is no guarantee that they are inside unit circle. A controller that produces very large (command) outputs for very small (error) inputs may be produced if a rapid close-loop response is requested (using $p \to 0$ in the z-plane) for slow plants (with $\sigma \to 0$ in the s-plane). Highly oscillatory controller responses may result if the roots of $\mathcal{A}_C(z)$ are near $z = -1$. Caution is required in the latter case because the continuous-time response of the plant may be highly oscillatory even though its discrete-time response looks reasonable at the sampling times, i.e. aliasing occurs. It should not be forgotten that a discrete-time controller is being used to control a continuous-time plant and that the plant model was discretized to support the design of the digital controller. It should also be noted that the discrete-time plant model is an exact solution for a perfect plant model connected to a zero-order hold and it is not an approximation. Thus the theoretical intra-sample response of the plant output may be examined by solving its linear state-space equations in continuous time using (12), for a given sequence of control commands and a specified initial state.

### Fourth-order polynomial design

In theory, the feedback system in Figure 1, has a zero steady-state error ($e$) for a step change in the reference ($r$) – because $L(z)$ already includes an integrator (i.e. a pole at $z = 1$) due to the plant model $\mathcal{H}_P(z)$. However, an additional integrator in $L(z)$ is needed to precisely track a ramp reference in azimuth, i.e. for a target with a constant azimuth rate, at steady sate. The required integrator is added to the controller $\mathcal{H}_C(z)$ by ensuring that $\mathcal{A}_C(z)$ has one root at $z = 1$. The procedure above is readily modified to satisfy this requirement.

The controller is augmented so that $\mathcal{H}_C(z)$ includes an integrator, i.e. a factor of $z/(z-1)$. This increases the order of the controller from three to four, thus for the second-order plant, the order of the closed-loop system is increased from five to six. With six poles at $p$, the denominator polynomial of $\mathcal{H}_F(z)$ is

$$\mathcal{A}_F(z) = (z-p)^6$$
$$= a_F[0]z^6 + a_F[1]z^5 + a_F[2]z^4 + a_F[3]z^3 + a_F[4]z^2 + a_F[5]z + a_F[6] \tag{29a}$$

where

$a_F[0] = 1, a_F[1] = -6p, a_F[2] = 15p^2, a_F[3] = -20p^3,$

$a_F[4] = 15p^4, a_F[5] = -6p^5$ and $a_F[6] = p^6$. (29b)

For the augmented controller we have

$$B_C(z) = \gamma_C(z-\beta_0)(z-\beta_1)(z-\beta_2)$$
$$= b_C[0]z^4 + b_C[1]z^3 + b_C[2]z^2 + b_C[3]z + b_C[4] \text{ with} \tag{30a}$$

$b_C[0] = 0$ and (30b)

$$\mathcal{A}_C(z) = z(z-1)(z-\alpha_0)(z-\alpha_1) \tag{30c}$$

$$= a_C[0]z^4 + a_C[1]z^3 + a_C[2]z^2 + a_C[3]z + a_C[4]$$

$$= z^2\{\bar{a}_C[0]z^2 + \bar{a}_C[1]z + \bar{a}_C[2]\} - z\{\bar{a}_C[0]z^2 + \bar{a}_C[1]z + \bar{a}_C[2]\} \text{ or} \tag{30d}$$

$$= \{\bar{a}_C[0]z^4 + \bar{a}_C[1]z^3 + \bar{a}_C[2]z^2\} - \{\bar{a}_C[0]z^3 + \bar{a}_C[1]z^2 + \bar{a}_C[2]z\} \text{ with} \tag{30e}$$

$$a_C[0] = \bar{a}_C[0] = 1 \tag{30f}$$

$$a_C[1] = \bar{a}_C[1] - \bar{a}_C[0] \tag{30g}$$

$$a_C[2] = \bar{a}_C[2] - \bar{a}_C[1] \tag{30h}$$

$$a_C[3] = -\bar{a}_C[2] \text{ and} \tag{30i}$$

$$a_C[4] = 0 \,. \tag{30j}$$

Note that the bar accents are used above to denote the unknown coefficients resulting from the $(z - \alpha_0)(z - \alpha_1)$ product. For the sum of polynomial products in (23), the following set of linear equations result:

$$
\begin{aligned}
a_F[0]z^6 &= a_P[0]z^2\bar{a}_C[0]z^4 \\
a_F[1]z^5 &= a_P[1]z^1\bar{a}_C[0]z^4 - a_P[0]z^2\bar{a}_C[0]z^3 \quad a_P[0]z^2\bar{a}_C[1]z^3 &&&& +b_P[0]z^2 b_C[1]z^3 \\
a_F[2]z^4 &= a_P[2]z^0\bar{a}_C[0]z^4 - a_P[1]z^1\bar{a}_C[0]z^3 \quad a_P[1]z^1\bar{a}_C[1]z^3 - a_P[0]z^2\bar{a}_C[1]z^2 \quad a_P[0]z^2\bar{a}_C[2]z^2 && +b_P[1]z^1 b_C[1]z^3 & +b_P[0]z^2 b_C[2]z^2 \\
a_F[3]z^3 &= \quad\quad -a_P[2]z^0\bar{a}_C[0]z^3 \quad a_P[2]z^0\bar{a}_C[1]z^3 - a_P[1]z^1\bar{a}_C[1]z^2 \quad a_P[1]z^1\bar{a}_C[2]z^2 - a_P[0]z^2\bar{a}_C[2]z^1 && +b_P[2]z^0 b_C[1]z^3 & +b_P[1]z^1 b_C[2]z^2 & +b_P[0]z^2 b_C[3]z^1 \\
a_F[4]z^2 &= \quad\quad\quad\quad -a_P[2]z^0\bar{a}_C[1]z^2 \quad a_P[2]z^0\bar{a}_C[2]z^2 - a_P[1]z^1\bar{a}_C[2]z^1 && +b_P[2]z^0 b_C[2]z^2 & +b_P[1]z^1 b_C[3]z^1 & +b_P[0]z^2 b_C[4]z^0 \\
a_F[5]z^1 &= \quad\quad\quad\quad\quad\quad -a_P[2]z^2\bar{a}_C[2]z^1 && +b_P[2]z^0 b_C[3]z^1 & +b_P[1]z^1 b_C[4]z^0 \\
a_F[6]z^0 &= && +b_P[2]z^0 b_C[4]z^0
\end{aligned}
$$

(31a)

After substituting, rearranging and simplifying:

$$
\begin{aligned}
a_F[1]z^5 - (a_P[1]z^5 - a_P[0]z^5) &= a_P[0]\bar{a}_C[1]z^5 \\
a_F[2]z^4 - (a_P[2]z^4 - a_P[1]z^4) &= a_P[1]\bar{a}_C[1]z^4 - a_P[0]\bar{a}_C[1]z^4 \quad a_P[0]\bar{a}_C[2]z^4 && +b_P[1] b_C[1]z^4 \\
a_F[3]z^3 + a_P[2]z^3 &= a_P[2]\bar{a}_C[1]z^3 - a_P[1]\bar{a}_C[1]z^3 \quad a_P[1]\bar{a}_C[2]z^3 - a_P[0]\bar{a}_C[2]z^3 && +b_P[2] b_C[1]z^3 & +b_P[1] b_C[2]z^3 \\
a_F[4]z^2 &= \quad -a_P[2]\bar{a}_C[1]z^2 \quad a_P[2]\bar{a}_C[2]z^2 - a_P[1]\bar{a}_C[2]z^2 && +b_P[2] b_C[2]z^2 & +b_P[1] b_C[3]z^2 \\
a_F[5]z^1 &= \quad\quad -a_P[2]\bar{a}_C[2]z^1 && +b_P[2] b_C[3]z^1 & +b_P[1] b_C[4]z^1 \\
a_F[6]z^0 &= && +b_P[2] b_C[4]z^0
\end{aligned}.
$$

(31b)

Or in matrix/vector form:

$$
\boldsymbol{v}_F = \begin{bmatrix} a_F[1] - a_P[1] + 1 \\ a_F[2] - a_P[2] + a_P[1] \\ a_F[3] + a_P[2] \\ a_F[4] \\ a_F[5] \\ a_F[6] \end{bmatrix},\; \boldsymbol{\chi}_P = \begin{bmatrix} 1 & 0 & 0 & 0 & 0 & 0 \\ a_P[1]-1 & 1 & b_P[1] & 0 & 0 & 0 \\ a_P[2]-a_P[1] & a_P[1]-1 & b_P[2] & b_P[1] & 0 & 0 \\ -a_P[2] & a_P[2]-a_P[1] & 0 & b_P[2] & b_P[1] & 0 \\ 0 & -a_P[2] & 0 & 0 & b_P[2] & b_P[1] \\ 0 & 0 & 0 & 0 & 0 & b_P[2] \end{bmatrix},\; \boldsymbol{\mu}_C = \begin{bmatrix} \bar{a}_C[1] \\ \bar{a}_C[2] \\ b_C[1] \\ b_C[2] \\ b_C[3] \\ b_C[4] \end{bmatrix}.
$$

(31c)

## Frequency design

The frequency method sacrifices agility, responsiveness and speed (i.e. 'adroitness') for resilience, reliability, and stability, when plant models are uncertain (i.e. 'robustness'); thus best-case errors are typically larger but worst-case errors are smaller. Transfer function representations $\mathcal{H}(z)$ are replaced by frequency-response representations $\mathcal{H}(\omega)$, where $\omega$ is the (relative) angular frequency in radians per sample ($\omega = \Omega/F_s$) and $\mathcal{H}(\omega)$ is determined by evaluating $\mathcal{H}(z)$ around the unit circle using $\mathcal{H}(\omega) = \mathcal{H}(z)|_{z=e^{i\omega}}$. Note that when the complex exponential $h(t) = e^{\sigma t + i\Omega t}$ is sampled using $t = T_s n$ and $\mathcal{Z}$-transformed, it has a pole at $z = re^{i\omega}$ where $r = e^{\sigma T_s}$ is the radial coordinate and $\omega = \Omega T_s$ is the angular coordinate in the complex z-plane. The approximate locations of dominant poles and zeros may be inferred from a frequency response. If a pole is at $\alpha$ in the complex z-plane then there is a peak in $|\mathcal{H}(\omega)|$ at the pole angle, i.e. at $\omega = \angle\alpha$, and the sharpness of the peak increases as $|\alpha| \to 1$, i.e. as the pole radius approaches the unit circle; where $|\blacksquare|$ and $\angle\blacksquare$ are the magnitude and angle operators for a complex argument. For example, a pole at $z = 0$ (i.e. a pure delay) affects all frequencies equally and a pole at $z = 1$ (i.e. an integrator) has a magnitude singularity at dc, i.e. $|\mathcal{H}(\omega)| = \infty$ at $\omega = 0$. Similarly, if a zero is at $\beta$ in the complex z-plane then there is a trough in $|\mathcal{H}(\omega)|$, i.e. at $\omega = \angle\beta$, and the sharpness and depth of the trough increases as $|\alpha| \to 1$, i.e. as the radius of the zero approaches the unit circle, for a perfect null, i.e. $|\mathcal{H}(\omega)| = 0$, when $|\alpha| = 1$. For the frequency design procedure, the complex z-plane (a mathematical abstraction) is replaced by the (tangible and measurable) frequency response; and bandwidth is used to specify the approximate duration of a system's transient response as an alternative to the pole radius (i.e. $r$).

The feedback system in Figure 1 requires negative feedback for input attenuation/cancellation and stability. For a sinusoidal signal, a change of sign corresponds to a phase lag (or lead) of $\pi$ radians which promotes decoherent disintegration for a *stable* feedback loop. A small delay shifts the phase of a low-frequency sinusoid by a small amount and the loop remains stable; however, a small delay shifts the phase of a high-frequency sinusoid by a larger amount. If the phase shift exceeds $\pm\pi$, and there is also amplification around the loop at that frequency, i.e. positive gain on a dB scale, then negative feedback effectively becomes positive feedback for coherent integration and an *unstable* feedback loop. Bode showed that all frequencies with a phase shift greater than or equal to $\pm\pi$ must be attenuated, for a feedback loop to be stable. Impulse, pulse, step, ramp, white noise, or coloured noise (etc.), inputs have energy spread over all frequencies, thus any finite input will induce an unbounded output if the frequency response $L(\omega)$ of the loop function $L(z)$ does not satisfy Bode's stability condition over all frequencies, where $L(\omega) = L(z)|_{z=e^{i\omega}}$.

If the output of $L(z)$ is negated and fed back into its input, then $L(\omega)$ determines whether the transfer function $\mathcal{H}_F(z)$ of the closed-loop system will be stable or unstable. For the system in Figure 1, $L(z) = \mathcal{H}_C(z)\mathcal{H}_P(z)$, as stated in (21). If the stability condition is not satisfied for $\mathcal{H}_P(\omega)$ then $\mathcal{H}_C(\omega)$ should be chosen to shape $L(\omega)$ so that stable feedback is attained – an essential design objective. Thus $\mathcal{H}_C(\omega)$ ensures that the magnitude of $L(\omega)$ is less than unity, wherever $L(\omega)$ has a phase shift of half a revolution or more. Furthermore, $\mathcal{H}_C(\omega)$ should shape the frequency response of $L(\omega)$ so that $\mathcal{H}_F(\omega)$ has sufficient bandwidth for a reasonable transient-response duration – a desirable design objective.

The so-called "sensitivity function" is defined as $S(z) = 1/\{1 + L(z)\}$. For the feedback system in Figure 1, the function that transfers the reference ($r$) to the controller input ($e$), i.e. $\mathcal{H}_F^{r \to e}(z)$ in (20a), and the function that transfers the disturbance ($d$) to the plant input ($x$), i.e. $\mathcal{H}_F^{d \to x}(z)$ in (20d), are both equal to $S(z)$. The sensitivity function also highlights the role that $L(z)$ plays in shaping the response of the closed-loop system. At steady state, for a sinusoidal input with a frequency of $\omega$, the tracking error (phase and magnitude) of a servomechanism and the disturbance attenuation of a regulator, are determined by $S(\omega)$, where $S(\omega) = S(z)|_{z=e^{i\omega}}$. Note that the magnitude of the sensitivity function is small, i.e. $|S(\omega)| \to 0$, at frequencies where the magnitude of the loop function is large, i.e. $|L(\omega)| \to \infty$, and that $|S(\omega)| \to 1$ as $|L(\omega)| \to 0$. Thus, $|L(\omega)|$ is an indication of a control loop's ability to influence plant behaviour (or exert control) as a function of frequency. Feedback control has no effect on sinusoidal ($r$ or $d$) inputs at frequencies where $|S(\omega)| = 1$; it attenuates ($e$ or $x$) outputs where $|S(\omega)| < 1$ and introduces distortion & bias where $|S(\omega)| \gg 1$.

A controller may satisfy Bode's condition for the (theoretical) plant model; however, differences in the magnitude and phase response for the actual plant may be sufficient to render the (operational) closed-loop system unstable. Robust controls are designed with ample stability margins to accommodate plant model uncertainty. Magnitude and/or phase margins may be used; however, the phase margin is used here because it is assumed that delays and latencies around the loop are the more likely form of model error.

For a second-order controller (e.g. PD or PI with a one-sample delay) the frequency-domain design procedure ensures that $\angle L(\widetilde{\omega}) = -\pi + \widetilde{\varphi}$ and $|L(\widetilde{\omega})| = 1$, where $\widetilde{\varphi}$ is the phase margin in radians and $\widetilde{\omega}$ is the bandwidth of the loop function, or the gain cross-over frequency, where the magnitude of $L(\omega)$, which is assumed to be a low-pass function, falls below unity [7]&[8]. For a third-order controller (e.g. PID with a one-sample delay) the extra degree of freedom is used to place a Nyquist null, i.e. $|L(\omega)| = 0$ at $\omega = \pi$, which is intended to further encourage magnitude roll-off at higher frequencies [8].

The controller is expressed as a linear combination of $K_\psi$ components multiplied by a pure one-sample delay to support the implementation of the feedback loop on a computer, i.e.

$\mathcal{H}_C(z) = \frac{1}{z}\sum_{k=0}^{K_\psi-1} c_k \psi_k(z)$ thus (32a)

$\mathcal{H}_C(\omega) = \frac{1}{z}\sum_{k=0}^{K_\psi-1} c_k \psi_k(\omega)$ where (32b)

$\mathcal{H}_C(\omega) = \mathcal{H}_C(z)|_{z=e^{i\omega}}$ (32c)

$\psi_k(\omega) = \psi_k(z)|_{z=e^{i\omega}}$ and (32d)

$c_k$ are the controller coefficients, to be determined. (32e)

The order $K_C$ of the controller depends on the number and order of the $\psi_k(z)$ components.

## Second-order frequency design

For a Proportional-Derivative (PD) controller we have $K_\psi = 2$ with

$\psi_0(z) = 1$ (the 'proportional' term) and (33a)

$\psi_1(z) = (z-1)/z$ (the 'derivative' term). (33b)

Expansion of the sum in (32a) yields the second-order transfer function ($K_C = K_\psi = 2$)

$\mathcal{H}_C(z) = \frac{1}{z}\{c_0 + c_1\,(z-1)/z\} = \frac{c_0 z + c_1(z-1)}{z^2} = \frac{(c_0+c_1)z - c_1}{z^2}$ (34a)

which has a Finite Impulse Response (FIR)

$\mathcal{H}_C(z) = \mathcal{B}_C(z)/\mathcal{A}_C(z)$ where (34b)

$$\mathcal{B}_C(z) = b_C[0]z^2 + b_C[1]z + b_C[2] \text{ with} \tag{34c}$$

$$b_C[0] = 0, b_C[1] = c_0 + c_1 \text{ and } b_C[2] = -c_1 \text{ and where} \tag{34d}$$

$$\mathcal{A}_C(z) = a_C[0]z^2 + a_C[1]z + a_C[2] \text{ with} \tag{34e}$$

$$a[0] = 1, a_C[1] = 0 \text{ and } a_C[2] = 0. \tag{34f}$$

The unknown controller coefficients $b_C[1]$ & $b_C[2]$, are determined by solving the simultaneous equations for $c_0$ & $c_1$:

$$L(-\widetilde{\omega}) = L_0(-\widetilde{\omega})c_0 + L_1(-\widetilde{\omega})c_1 \text{ and} \tag{35a}$$

$$L(+\widetilde{\omega}) = L_0(+\widetilde{\omega})c_0 + L_1(+\widetilde{\omega})c_1 \text{ where} \tag{35b}$$

$$L(+\widetilde{\omega}) = e^{+i(\widetilde{\varphi}-\pi)} \tag{35c}$$

is the desired frequency response of the loop function at $+\widetilde{\omega}$

$$L(-\widetilde{\omega}) = L^*(+\widetilde{\omega}) = e^{-i(\widetilde{\varphi}-\pi)} \tag{35d}$$

is the desired frequency response of the loop function at $-\widetilde{\omega}$
(note that ■* is the complex conjugation operator)

$$L_k(\widetilde{\omega}) = L_k(z)|_{z=e^{i\widetilde{\omega}}} \text{ with} \tag{35e}$$

$$L_k(z) = \frac{1}{z}\mathcal{H}_P(z)\psi_k(z) \text{ and where} \tag{35f}$$

$\mathcal{H}_P(z)$ for the nominal plant model is defined in (19).

Or in matrix/vector notation

$$\boldsymbol{l} = \boldsymbol{L}\boldsymbol{c} \text{ which is readily solved using} \tag{36a}$$

$$\boldsymbol{c} = \boldsymbol{L}^{-1}\boldsymbol{l} \text{ where} \tag{36b}$$

$$\boldsymbol{l} = \begin{bmatrix} e^{-i(\widetilde{\varphi}-\pi)} \\ e^{+i(\widetilde{\varphi}-\pi)} \end{bmatrix} \tag{36d}$$

$$\boldsymbol{L} = \begin{bmatrix} L_0(-\widetilde{\omega}) & L_1(-\widetilde{\omega}) \\ L_0(+\widetilde{\omega}) & L_1(+\widetilde{\omega}) \end{bmatrix} \text{ and} \tag{36e}$$

$$\boldsymbol{c} = \begin{bmatrix} c_0 \\ c_1 \end{bmatrix}. \tag{36f}$$

### Third-order frequency design

At steady state, the controller requires an integrator to track a ramp reference ($r$) with a zero-error input ($e$) to the controller or to supress a ramp disturbance ($d$) for a zero-sum input ($x$) to the plant. For this Proportional-Integral-Derivative (PID) controller we have $K_\psi = 3$ with

$$\psi_0(z) = 1 \text{ (the 'proportional' term)} \tag{37a}$$

$$\psi_1(z) = (z-1)/z \text{ (the 'derivative' term) and} \tag{37b}$$

$$\psi_2(z) = z/(z-1) \text{ (the 'integral' term).} \tag{37c}$$

Expansion of the sum in (32a) for this $K_c = K_\psi = 3$ controller (with delay included) yields the third-order transfer function

$$\mathcal{H}_C(z) = \tfrac{1}{z}\{c_0 + c_1\,(z-1)/z + c_1\,z/(z-1)\} \tag{38a}$$

which has an Infinite Impulse Response (IIR)

$$\mathcal{H}_C(z) = \mathcal{B}_C(z)/\mathcal{A}_C(z) \text{ where} \tag{38b}$$

$$\mathcal{B}_C(z) = b_C[0]z^3 + b_C[1]z^2 + b_C[2]z + b_C[3] \text{ with} \tag{38c}$$

$$b_C[0] = 0,\ b_C[1] = c_0 + c_1 + c_2,\ b_C[2] = -c_0 - 2c_1 \text{ and } b_C[3] = c_1 \text{ and where} \tag{38d}$$

$$\mathcal{A}_C(z) = a_C[0]z^3 + a_C[1]z^2 + a_C[2]z + a_C[3] \text{ with} \tag{38e}$$

$$a[0] = 1,\ a_C[1] = -1,\ a_C[2] = 0 \text{ and } a_C[3] = 0. \tag{38f}$$

For the complex frequency-response of a transfer function with a real impulse-response, two degrees of freedom are 'consumed' to satisfy the phase-margin constraint. The third degree of freedom is used to lower the high-frequency noise gain of the loop. Excessive control action increases system operation and maintenance costs, e.g. due to greater power consumption or mechanical wear-and-tear; furthermore, if derivative control is not used judiciously, it is notorious for producing 'jittery' or 'hyperactive' controls that respond emphatically and frequently to insignificant inputs. The placement of a 'Nyquist null' in $L(\omega)$ addresses these potential problems by attenuating high-frequency noise and it supports the phase-margin constraint because it encourages magnitude roll-off at high frequencies, for closed-loop stability. For the simultaneous equations in (36a) we therefore have

$$\boldsymbol{l} = \begin{bmatrix} e^{-i(\tilde{\varphi}-\pi)} \\ e^{+i(\tilde{\varphi}-\pi)} \\ 0 \end{bmatrix} \tag{39a}$$

$$\boldsymbol{L} = \begin{bmatrix} L_0(-\tilde{\omega}) & L_1(-\tilde{\omega}) & L_2(-\tilde{\omega}) \\ L_0(+\tilde{\omega}) & L_1(+\tilde{\omega}) & L_2(+\tilde{\omega}) \\ L_0(\pi) & L_1(\pi) & L_2(\pi) \end{bmatrix} \text{ and} \tag{39b}$$

$$\boldsymbol{c} = \begin{bmatrix} c_0 \\ c_1 \\ c_2 \end{bmatrix}. \tag{39c}$$

## Controller realization and simulation

Linear state-space representations were used above to model plant dynamics in continuous time. They are also useful for realizing the controller in a digital computer and for simulating the feedback loop in discrete time [3]. Furthermore, they provide an alternative way of designing controls by pole placement that effectively 'automates' the 'manual' polynomial manipulations and equation definitions in (27) & (31); however, this state-space design procedure (see [3] for details) is not discussed here.

A discrete-time transfer function $\mathcal{H}(z) = \mathcal{B}(z)/\mathcal{A}(z)$ with coefficients $b[k]$ and $a[k]$ for $k = 0 \ldots K-1$ (with $a[0] = 1$) has the following state-space representation:

$$\boldsymbol{G} = \begin{bmatrix} -a[1] & -a[2] & \cdots & -a[K-2] & -a[K-1] & -a[K-0] \\ 1 & 0 & \cdots & 0 & 0 & 0 \\ 0 & 1 & \cdots & 0 & 0 & 0 \\ \vdots & \vdots & \ddots & \vdots & \vdots & \vdots \\ 0 & 0 & \cdots & 1 & 0 & 0 \\ 0 & 0 & \cdots & 0 & 1 & 0 \end{bmatrix}_{K \times K} \tag{40a}$$

$$\boldsymbol{H} = \begin{bmatrix} 1 \\ 0 \\ \vdots \\ 0 \\ 0 \\ 0 \end{bmatrix}_{K \times 1} \tag{40b}$$

$$\boldsymbol{C}^{\mathrm{T}} = \begin{bmatrix} b[1] - b[0]a[1] \\ b[2] - b[0]a[2] \\ \vdots \\ b[K-2] - b[0]a[K-2] \\ b[K-1] - b[0]a[K-1] \\ b[K-0] - b[0]a[K-0] \end{bmatrix}_{K \times 1} \tag{40c}$$

$$\boldsymbol{D} = b[0] \,. \tag{40d}$$

In this so-called "canonical" form, the internal states of the system have no physical significance. They are simply the contents of a sequence of delay registers that are used to compute the output of the system for a given input. Note that the discrete-time transfer functions in this document are expressed as a ratio of polynomials with positive powers of $z$, which makes it easier to quickly identify the locations of poles and zeros; however, they are realized using delays, which are all non-positive powers of $z$. The former representation is simply converted to the latter, by dividing the numerator and denominator polynomials by the highest power of $z$.

The above (polynomial and frequency) controls were designed with a pure one-sample delay incorporated, i.e. a factor of $z^{-1}$. When the controller is realized on a digital computer and connected to the external environment, an infinite outer loop is used to manage input/output operations and error-signal evaluation. It applies the one-sample delay to the error signal ($e$) and closes the loop to complete one feedback cycle. Thus, $z^{-1}$ is factored out of $\mathcal{H}_C(z)$ and a linear state-space representation of the *modified* controller $\mathcal{H}_{\hat{C}}(z)$ is applied inside this loop, i.e.

$$\mathcal{H}_C(z) = z^{-1}\mathcal{H}_{\hat{C}}(z) \text{ thus} \tag{41a}$$

$$K_{\hat{C}} = K_C - 1 \,. \tag{41b}$$

Pseudocode to realize and simulate the feedback loop is provided below.

*Initialize loop variables:*

$\boldsymbol{w}_{\hat{C}} = \boldsymbol{0}_{K_{\hat{C}} \times 1}$
$\boldsymbol{w}_P = \boldsymbol{0}_{K_P \times 1}$
$\hat{e} = 0$
For $n = 0 \ldots \infty$:
    *Compute $u[n]$:*
    $x_{\hat{C}} = \hat{e}$
    $\widehat{\boldsymbol{w}}_{\hat{C}} = \boldsymbol{G}_{\hat{C}} \boldsymbol{w}_{\hat{C}} + \boldsymbol{H}_{\hat{C}} x_{\hat{C}}$
    $y_{\hat{C}} = \boldsymbol{C}_{\hat{C}} \boldsymbol{w}_{\hat{C}} + \boldsymbol{D}_{\hat{C}} x_{\hat{C}}$
    $\boldsymbol{w}_{\hat{C}} = \widehat{\boldsymbol{w}}_{\hat{C}}$
    $u[n] = y_{\hat{C}}$
    *Compute $y[n]$:*
    $x[n] = u[n] + d[n]$
    $x_P = x[n]$
    $\widehat{\boldsymbol{w}}_P = \boldsymbol{G}_P \boldsymbol{w}_P + \boldsymbol{H}_P x_P$
    $y_P = \boldsymbol{C}_P \boldsymbol{w}_P + \boldsymbol{D}_P x_P$
    $\boldsymbol{w}_P = \widehat{\boldsymbol{w}}_P$
    $y[n] = y_P$
    *Wait here until $t = T_s n$, then continue:*
    *Input $y[n]$:*
    *Output $u[n]$:*
    $e[n] = r[n] - y[n]$
    $\hat{e} = e[n]$
End for $n$     (42)

In (42) $\boldsymbol{0}_{M \times N}$ is an $M \times N$ matrix of zeros and the $\blacksquare_{\hat{C}}$ & $\blacksquare_P$ subscripts pertain to the modified controller (i.e. without the delay) and the plant, respectively. The grey statements denote simulation operations where the plant is unavailable. In a generic online realization, $u[n]$ is output to the plant (plus actuator) via a digital-to-analogue converter and $y[n]$ is input from the plant (plus sensor) via an analogue-to-digital converter, for the computation of $e[n]$; however, the error signal is computed from the $n$th image frame in this tracking camera example.

## Discussion

### Plant parameterization

The controls in the next section are all designed for an assumed or *nominal* second-order plant model with a pole at $s = 0$ (due to the first-order integrating term) and a real pole at $s = -10$, i.e. $\sigma = -10$ and $\Omega = 0$ (due to the first-order lag term). The response of the plant, for various values of $\sigma$ are shown in Figure 2. The robustness of the controls designed for this nominal plant model is then checked using a *perturbed* plant model with an additional delay of one sample that is not considered in the controller design.

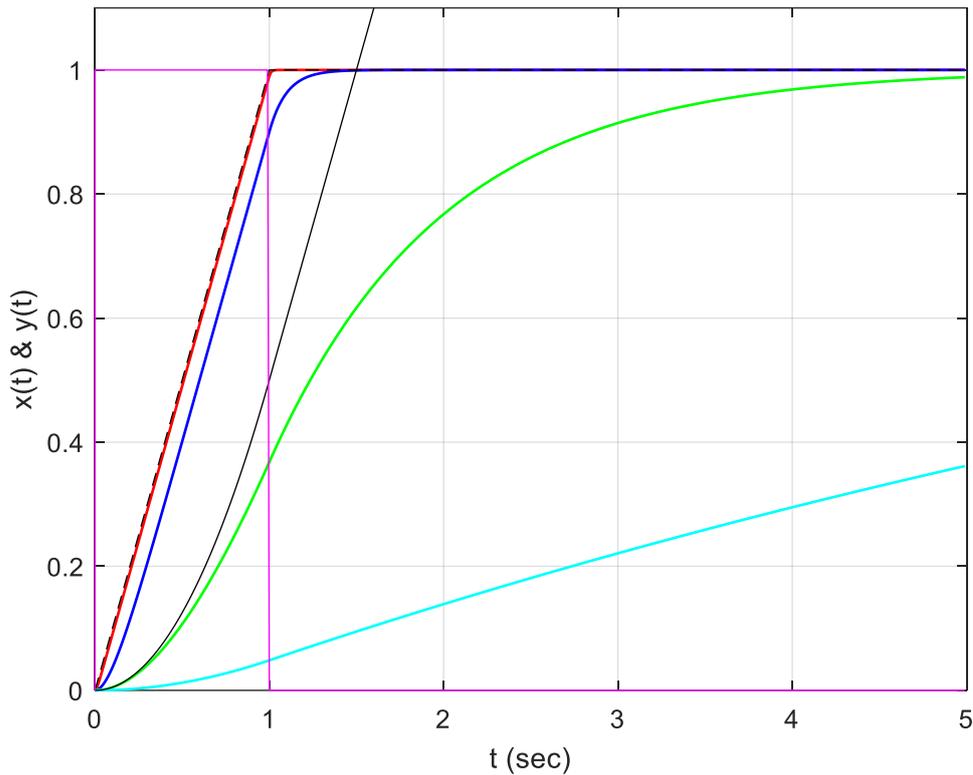

*Figure 2. Response $y(t)$ of a second-order plant model with a pole at $s = 0$ and at $s = \sigma$, for $\sigma = -100$ (red line), $\sigma = -10$ (blue line), $\sigma = -1$ (green line), and $\sigma = -0.1$ (cyan line), for a unit pulse input $x(t)$ (magenta line). Responses of an integrator (dashed black line) and a double integrator (solid black line) are also shown.*

## Controller analysis

Adroit *and* robust control is the always the objective; however, one is achieved at the expense of the other. Adroit (also referred to as "high-performance" [5]) controls quickly adjust to input changes and rapidly drive internal signals towards zero, whereas robust controls maintain closed-loop stability for larger plant-model errors. The sensitivity function quantifies both the adroitness and the robustness of a feedback system [5].

The frequency response of the function that transfers the reference ($r$) to the tracking error signal ($e$) and the disturbance ($d$) to the net plant input ($x$), i.e. (20a) & (20d), is equal to the sensitivity function. Thus for adroit control, $|S(\omega)|$ should be small over the expected range of input frequencies, which is typically near dc. Furthermore, for a $K$th-order monomial input (i.e. $t^K$, for $t \geq 0$) the first $K + 1$ derivatives of $S(\omega)$ (with respect to frequency) should equal zero at dc (i.e. $\omega = 0$), for $e$ and $x$ to converge on zero at steady state (highly desirable).

Nyquist plots are an alternative graphical representation of a system's frequency response $H(\omega)$ that hide the frequency axis. They are polar plots of $|H(\omega)|$ and $\angle H(\omega)$ and for low-pass systems the Nyquist curve spirals in towards zero as the magnitude rolls off at high frequencies. The robustness of a stable closed-loop system is readily gleaned from the Nyquist plot of $L(\omega)$, i.e. the frequency response of the loop function. Magnitude and phase perturbations from the nominal plant shift the Nyquist curve and when the curve passes over the critical point at $-1$ (corresponding to a phase of $\pm\pi$ and a magnitude of 1) the feedback loop becomes unstable (for any input). Thus the shortest distance from the curve to the critical point is an indication of the smallest model perturbation, in magnitude and phase, applied at any frequency, that causes instability. The distance from the critical point to any point of the curve is equal to $|1 + L(\omega)|$, which is equal to the dominator of $|S(\omega)|$, thus the distance is equal to $1/|S(\omega)|$ and the maximum of $|S(\omega)|$ corresponds to the shortest distance; therefore, robust controls have a small sensitivity peak [5],[6]&[9].

Bode showed that the integral (over frequency) of the natural log of the sensitivity function's magnitude, for a stable system, is zero. For adroit controls, $|S(\omega)|$ is small over a wide frequency range and for robust controls, peak $|S(\omega)|$ is small; however, Bode's integral indicates that decreasing either must increase the other. Reducing sensitivity magnitude near dc (for improved adroitness) simply 'transfers it' outside the effective bandwidth of the loop (for decreased robustness), like a pile of sand next to a hole created hastily by a digger [9].

The polynomial method computes the controller coefficients from the desired adroitness (i.e. the closed-loop pole positions) whereas the frequency method computes the controller coefficients from the desired robustness (i.e. the phase margin) *and* the desired adroitness (i.e. the bandwidth, which is set via the gain cross-over frequency); however, a stable frequency solution may not be possible for the specified controller poles. If the solution reached via the polynomial method has inadequate robustness (as determined from its sensitivity function peak) then different closed-loop poles should be considered. If the solution reached via the frequency method is unstable (as determined by the poles of the closed-loop system) then different controller poles should be used.

Robust stability is typically a high priority and it is readily quantified using the peak of the sensitivity function, which should be as close to unity as possible. For the system in Figure 1, $\mathcal{H}_F^{r \to e}(z)$ and $\mathcal{H}_F^{d \to x}(z)$ are equal to $S(z)$, thus it's reference-tracking performance and disturbance-rejection behaviour are the same. For the system in Figure 1, the 'width' of the sensitivity function's low-frequency passband or the inverse 'width' of its impulse response (which are equivalent) are an indication of adroitness. However, peaks and bandwidths of other transfer functions in (20) may be more relevant in other systems with other requirements [5]&[6].

In all feedback systems the sensitivity function is important. In addition to quantifying robustness, it also determines the response of the error signal ($e$) to references ($r$) and the response of the plant input ($x$) to disturbances ($d$). In some problems there may be limits on what can be applied to the plant (via $u$), e.g. when actuators are weak and fuel is finite, and in these cases, $\mathcal{H}_F^{r \to u}(z)$ & $\mathcal{H}_F^{d \to u}(z)$ are also of interest. In other problems there may be constraints on what the plant is permitted to output, e.g. to prevent mechanical or electrical degradation & failure, and in these cases $\mathcal{H}_F^{r \to y}(z)$ & $\mathcal{H}_F^{d \to y}(z)$ would be of interest. For the system in Figure 1, $\mathcal{H}_F^{r \to y}(z) = -\mathcal{H}_F^{d \to u}(z) = T(z)$, where $T(z)$ is the so-called "complementary sensitivity function" because $T(z) = 1 - S(z)$.

These tuning and analysis principles are depicted and summarized below for third-order polynomial (see Figure 3 - Figure 9 & Table 1), fourth-order polynomial (see Figure 10 - Figure 16 & Table 2), second-order frequency (see Figure 17 - Figure 23 & Table 3), and the third-order frequency (see Figure 24 - Figure 30 & Table 4), designs. All designs incorporate an additional pure delay to realize the digital feedback loop; the fourth-order polynomial and third-order frequency designs also incorporate an integrator (for improved steady-state reference tracking). The other single pole of the frequency design is at zero (for a standard proportional-derivative structure); whereas the other two poles of the polynomial design appear anywhere in the z-plane as a complex-conjugate pair.

### Analysis of polynomial designs

The pole-zero map in the complex z-plane for the polynomial designs indicate that an unstable controller is required for the more extreme configurations, i.e. the $p = 0.7$ tuning for the third-order design (see Figure 3) and the $p = 0.2$ tuning for the fourth-order design (see Figure 10). These configurations correspond to closed-loop responses with narrow and wide bandwidths respectively. Pole-zeros maps for the closed-loop system are not provided for polynomial designs because all poles are placed (approximately) at $p$ (see $r_0$ in Table 1 & Table 2 and note that $r_0 \cong p$ for all tunings). Stable closed-loop systems (with $r_0 < 1$) that are formed using unstable controls should be avoided in practice if possible [6]. As the complex poles of the controller approach the unit circle, tall and narrow peaks in $|L(\omega)|$ are formed that indicate the presence of resonant modes (see the upper subplots of Figure 5 & Figure 12). For both polynomial designs, there is a magnitude singularity at dc due to one pole on the unit circle (at $z = 1$) introduced by the plant for the third-order design and two poles on the unit circle from the plant and the controller for the fourth-order design.

Polynomial designs with a wide bandwidth (using $p \to 0$) are adroit but not robust; whereas polynomial designs with a narrow bandwidth (using $p \to 1$) are robust but not adroit. Clearly, placing the closed-loop poles well within the unit circle (i.e. near zero) does not guarantee robustness; furthermore, introducing an integrator degrades the stability margins significantly. For digital control systems, the delay margin $\tilde{\Delta}$ (in samples) is arguably a more meaningful measure of robustness than the phase margin $\tilde{\varphi}$ (in radians), where $\tilde{\Delta} = \tilde{\varphi}/\tilde{\omega}$ or $\tilde{\Delta} = \tilde{\varphi}/2\pi\tilde{f}$, where $\tilde{\omega}$ and $\tilde{f}$ are the gain cross-over frequency or desired bandwidth (in radians per sample or cycles per sample, respectively). The delay margin indicates that greater phase margins are required for wider bandwidths, for the same degree of robustness with respect to unmodelled delays. The delay margins of the third-order polynomial designs (see Table 1) are adequate and improve substantially as $p$ is increased (for a narrower closed-loop bandwidth); however, the delay margins of the fourth-order designs (see Table 2) are meagre. Less-than-unity delay margins indicate that delays (i.e. phase-only perturbations) of only one sample are sufficient to render the feedback loop unstable. These perturbations are generally assumed to occur in the plant, but in digital systems they may occur anywhere (and randomly) around the loop. The step responses (see Figure 9 & Figure 16) and the maximum closed-loop pole radius (see $r_0$ in Table 1 & Table 2) for a perturbed plant with an additional (unmodelled) one-sample delay, that is not considered in controller design, validate and illustrate the delay margin as a measure of robustness.

*Third-order configuration via polynomial design*

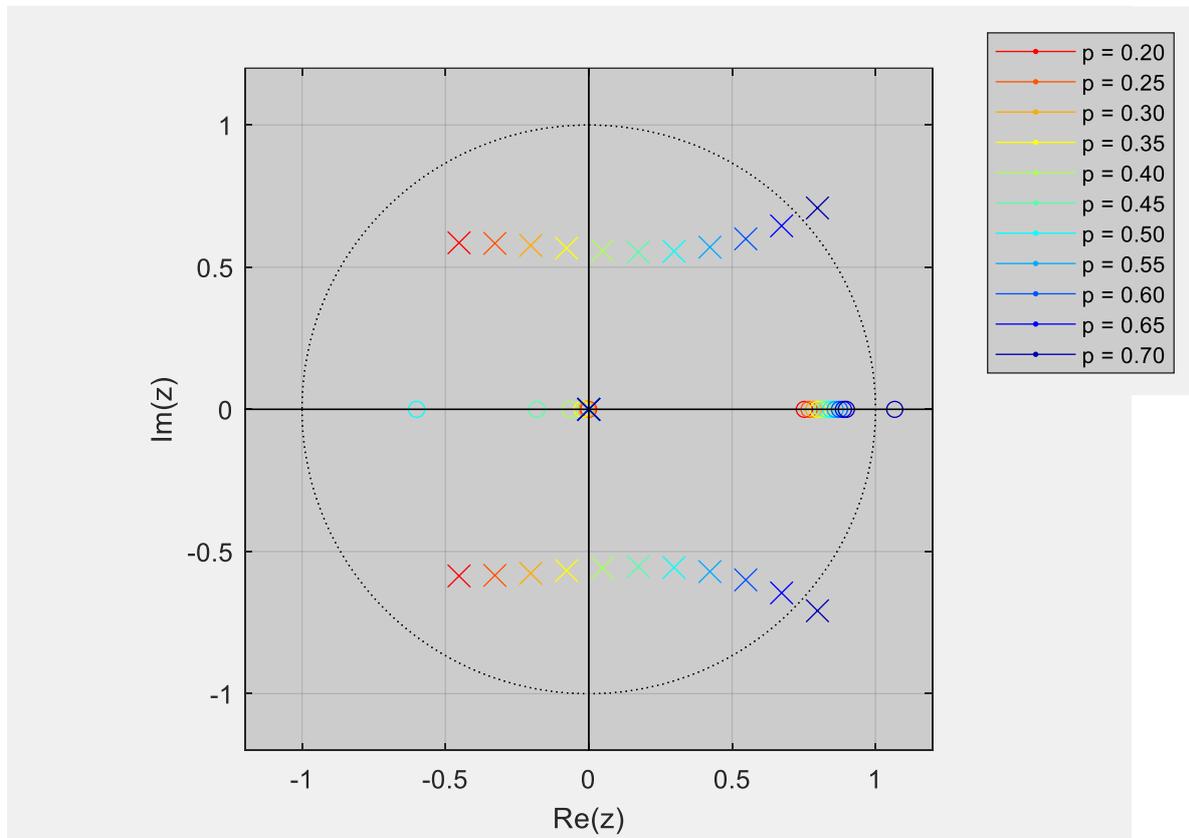

*Figure 3. Third-order controller designed via the polynomial method using a variety of different closed-loop pole positions (see legend). Positions of the poles ('X' tokens) and zeros ('O' tokens) of the controller transfer function in the complex z-plane.*

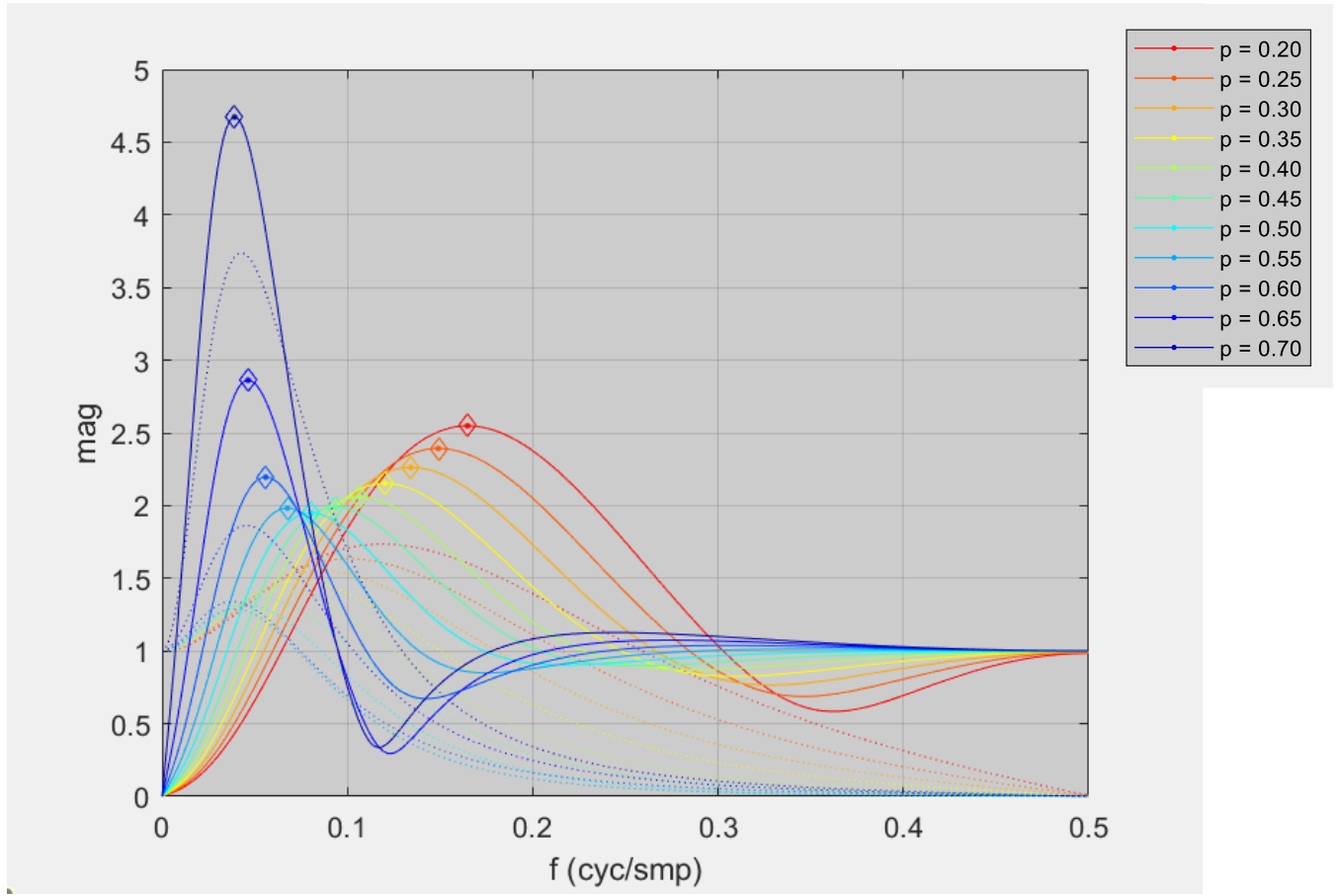

*Figure 4. Third-order controller designed via the polynomial method using a variety of different closed-loop pole positions (see legend). Magnitude of sensitivity function (solid lines), sensitivity maxima (diamond tokens) and complementary sensitivity function (dotted lines).*

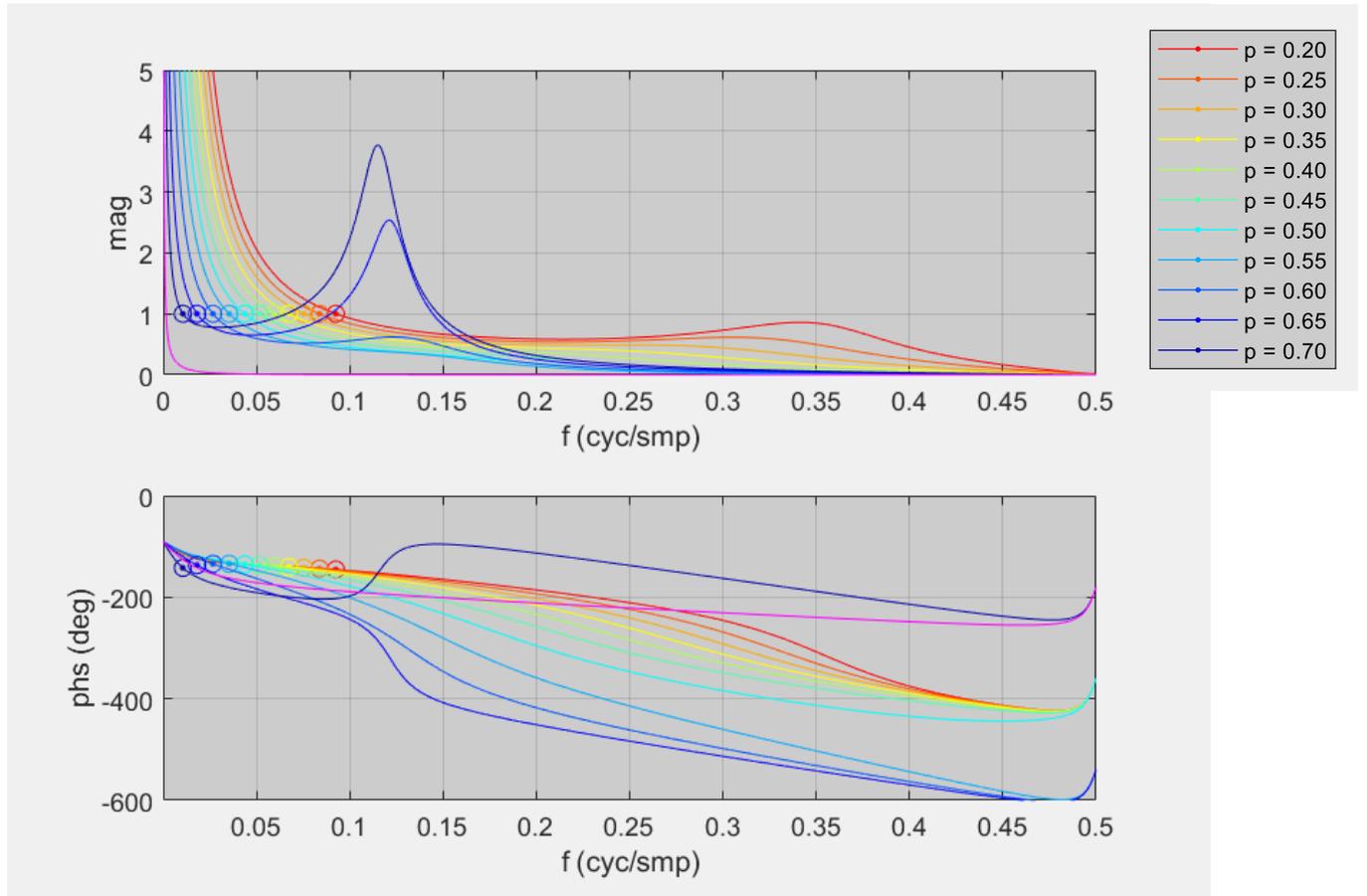

*Figure 5. Third-order controller designed via the polynomial method using a variety of different closed-loop pole positions (see legend). Frequency response of loop function (prismatic lines) and plant model (magenta line); magnitude (upper subplot) and phase (lower subplot). Magnitude and phase at gain cross-over frequency of loop function are also shown (circle tokens).*

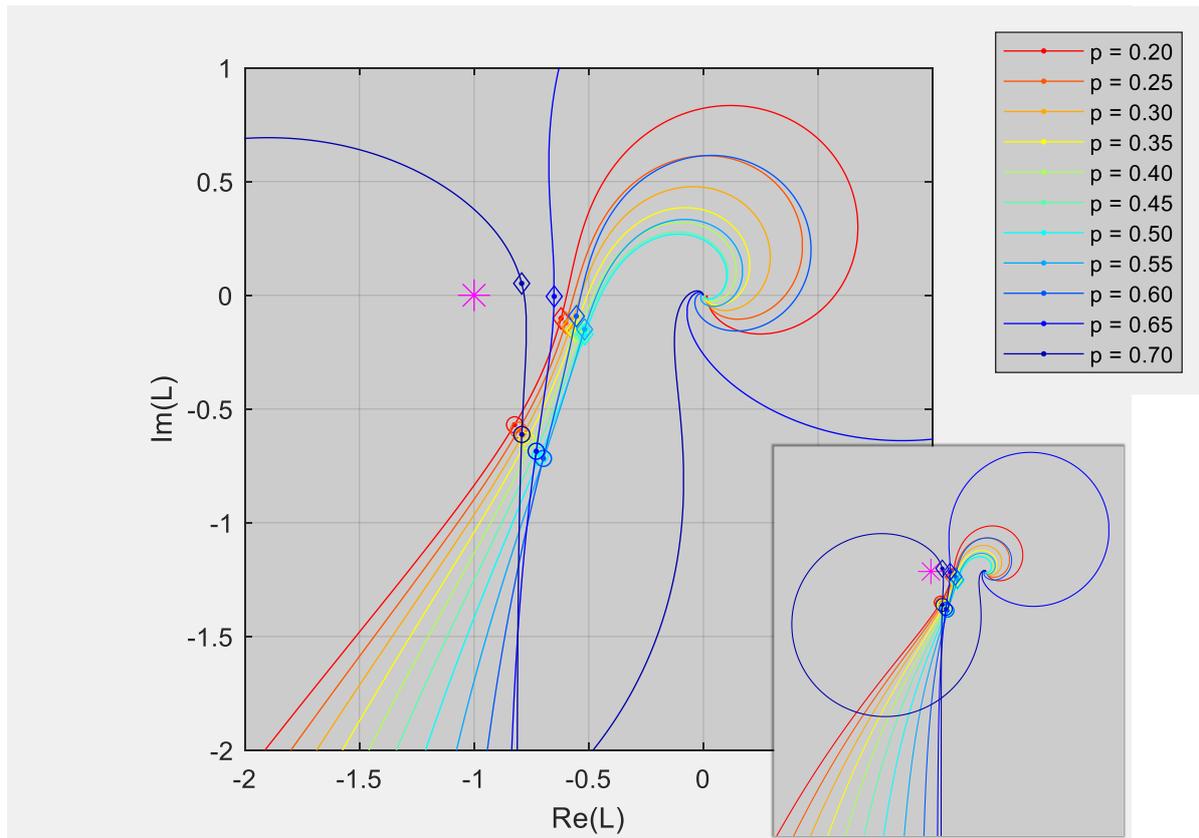

*Figure 6. Third-order controller designed via the polynomial method using a variety of different closed-loop pole positions (see legend). Nyquist plots for loop function with phase and magnitude at the gain cross-over frequency (circle tokens) and at the maximum sensitivity frequency (diamond tokens). The critical point at which the magnitude is unity and the phase shift is $\pm\pi$ is also shown (magenta asterisk). Inset shows a wider view and an encirclement of the critical point due to the unstable controller for $p = 0.7$.*

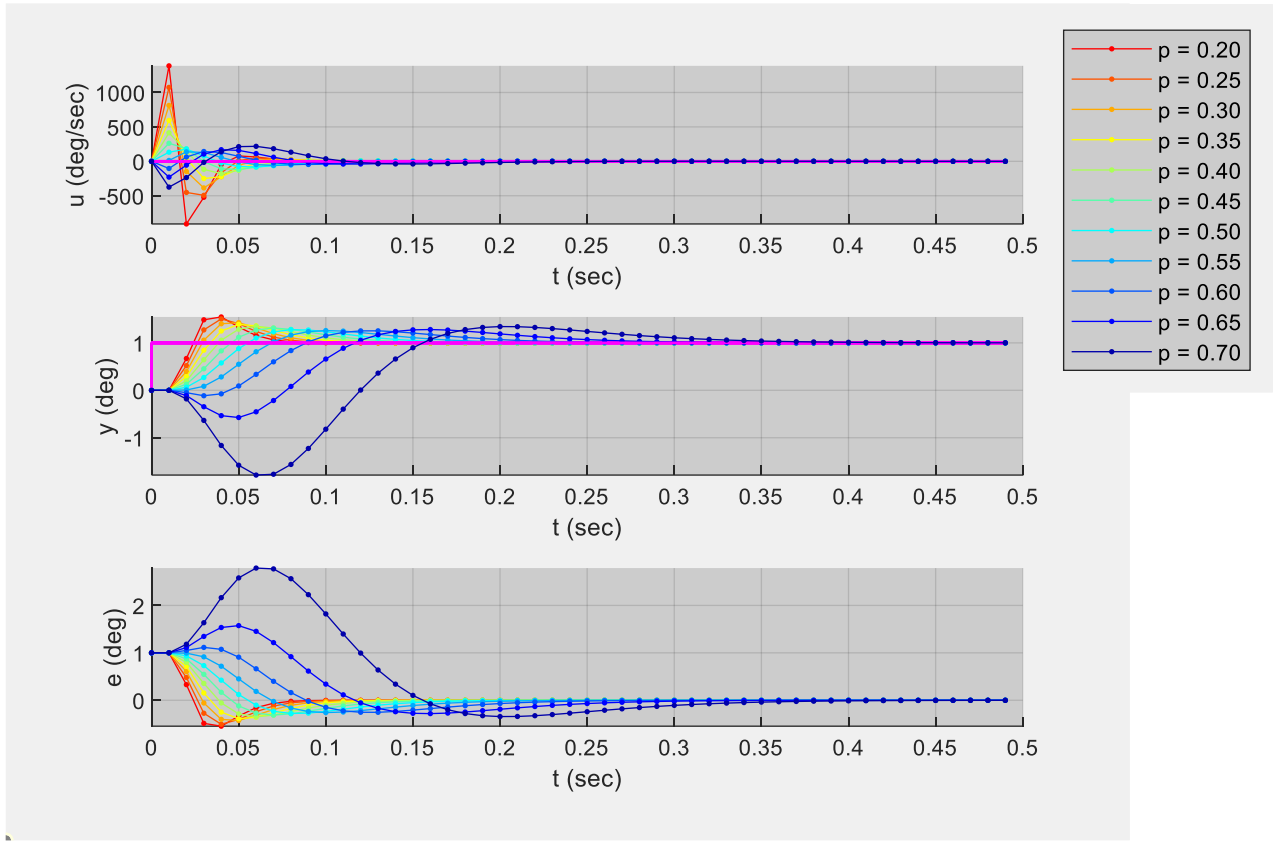

*Figure 7. Third-order controller designed via the polynomial method using a variety of different closed-loop pole positions (see legend). Closed-loop response for the nominal plant. Command signal (u, top subplot), plant output (y, middle subplot) and error signal (e, bottom subplot) for a unit step reference (r, magenta line, in middle subplot) and zero disturbances ($d = 0$, magenta line, in top subplot).*

https://arxiv.org/abs/2211.09932

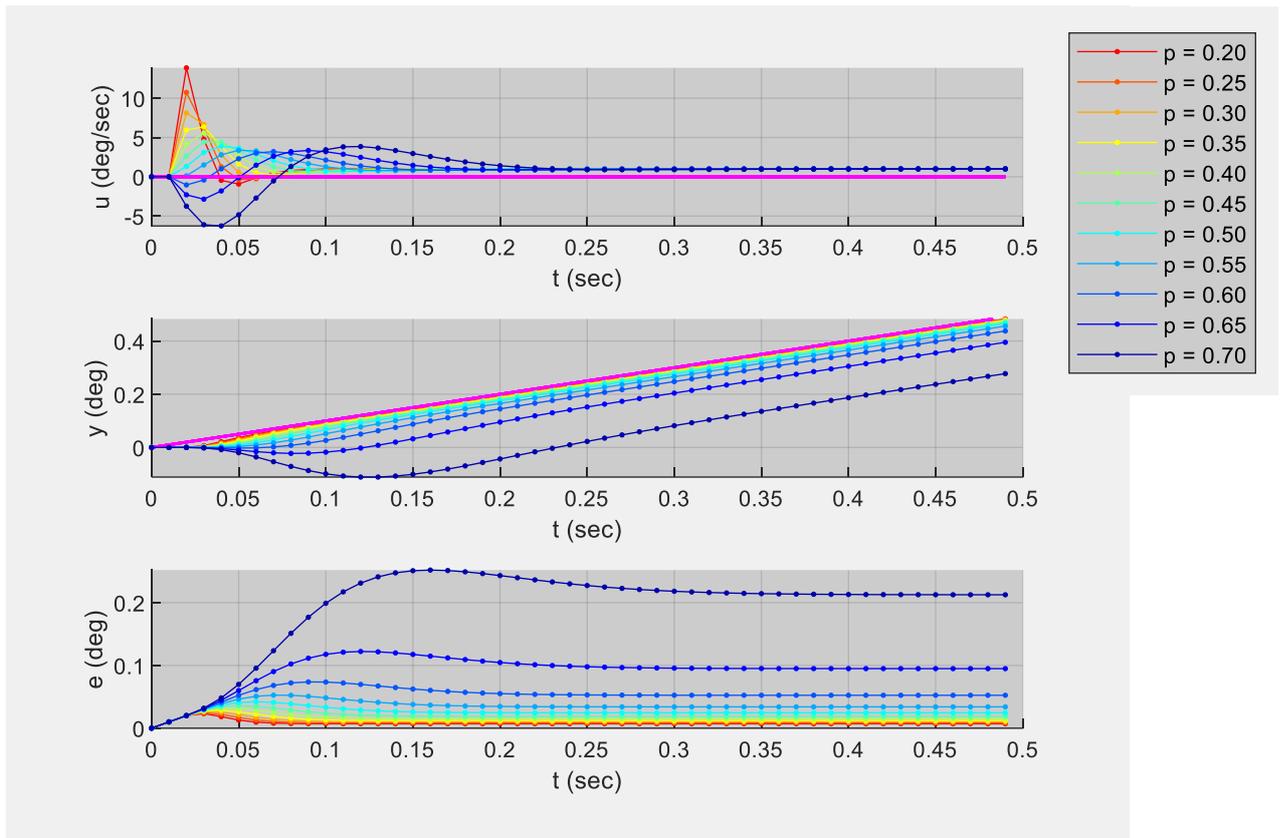

*Figure 8. Third-order controller designed via the polynomial method using a variety of different closed-loop pole positions (see legend). Closed-loop response for the nominal plant. Command signal (u, top subplot), plant output (y, middle subplot) and error signal (e, bottom subplot) for a ramp reference (r, magenta line, in middle subplot) and zero disturbances ($d = 0$, magenta line, in top subplot).*

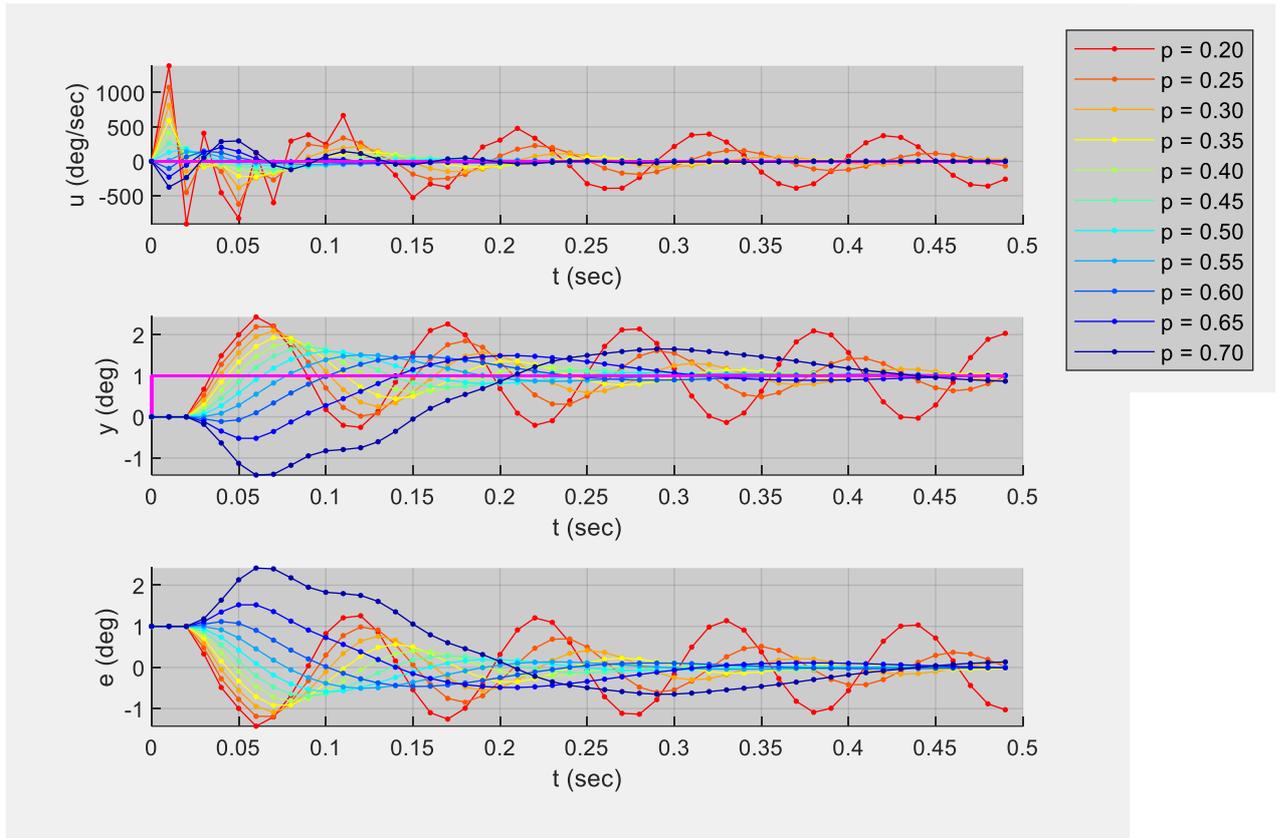

*Figure 9. Third-order controller designed via the polynomial method using a variety of different closed-loop pole positions (see legend). Closed-loop response for the perturbed plant with an unmodelled one-sample delay. Command signal (u, top subplot), plant output (y, middle subplot) and error signal (e, bottom subplot) for a unit step reference (r, magenta line, in middle subplot) and zero disturbances ($d = 0$, magenta line, in top subplot).*

*Table 1. Robustness of the third-order controller designed via the polynomial method. Phase margin ($\tilde{\varphi}$, deg), delay margin ($\tilde{\Delta}$, smp), and maximum closed-loop pole radius ($r_0$), for the nominal plant model, as a function of closed-loop pole position (p). Maximum closed-loop pole radius for perturbed plant model used in Figure 9 is also shown ($r_1$).*

| $p$ | $\tilde{\varphi}$ | $\tilde{\Delta}$ | $r_0$ | $r_1$ |
|---|---|---|---|---|
| 0.20 | 34.65 | 1.04 | 0.200 | 0.994 |
| 0.25 | 36.33 | 1.20 | 0.250 | 0.971 |
| 0.30 | 37.99 | 1.40 | 0.301 | 0.950 |
| 0.35 | 39.67 | 1.64 | 0.351 | 0.932 |
| 0.40 | 41.37 | 1.93 | 0.401 | 0.916 |
| 0.45 | 43.08 | 2.32 | 0.450 | 0.903 |
| 0.50 | 44.69 | 2.84 | 0.501 | 0.896 |
| 0.55 | 45.86 | 3.60 | 0.551 | 0.895 |
| 0.60 | 45.76 | 4.76 | 0.601 | 0.904 |
| 0.65 | 43.24 | 6.62 | 0.651 | 0.920 |
| 0.70 | 37.71 | 9.90 | 0.701 | 0.941 |

*Fourth-order configuration via polynomial design*

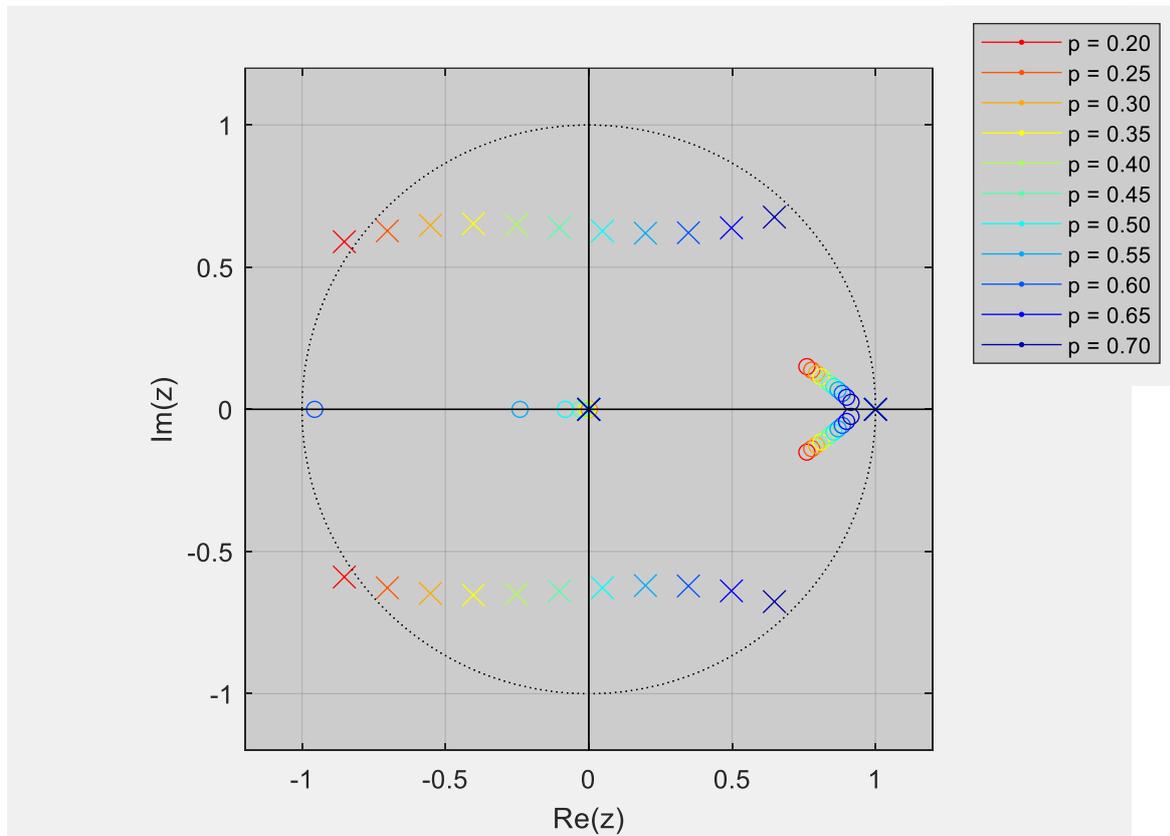

*Figure 10. Fourth-order controller designed via the polynomial method using a variety of different closed-loop pole positions (see legend). Positions of the poles ('X' tokens) and zeros ('O' tokens) of the controller transfer function in the complex z-plane.*

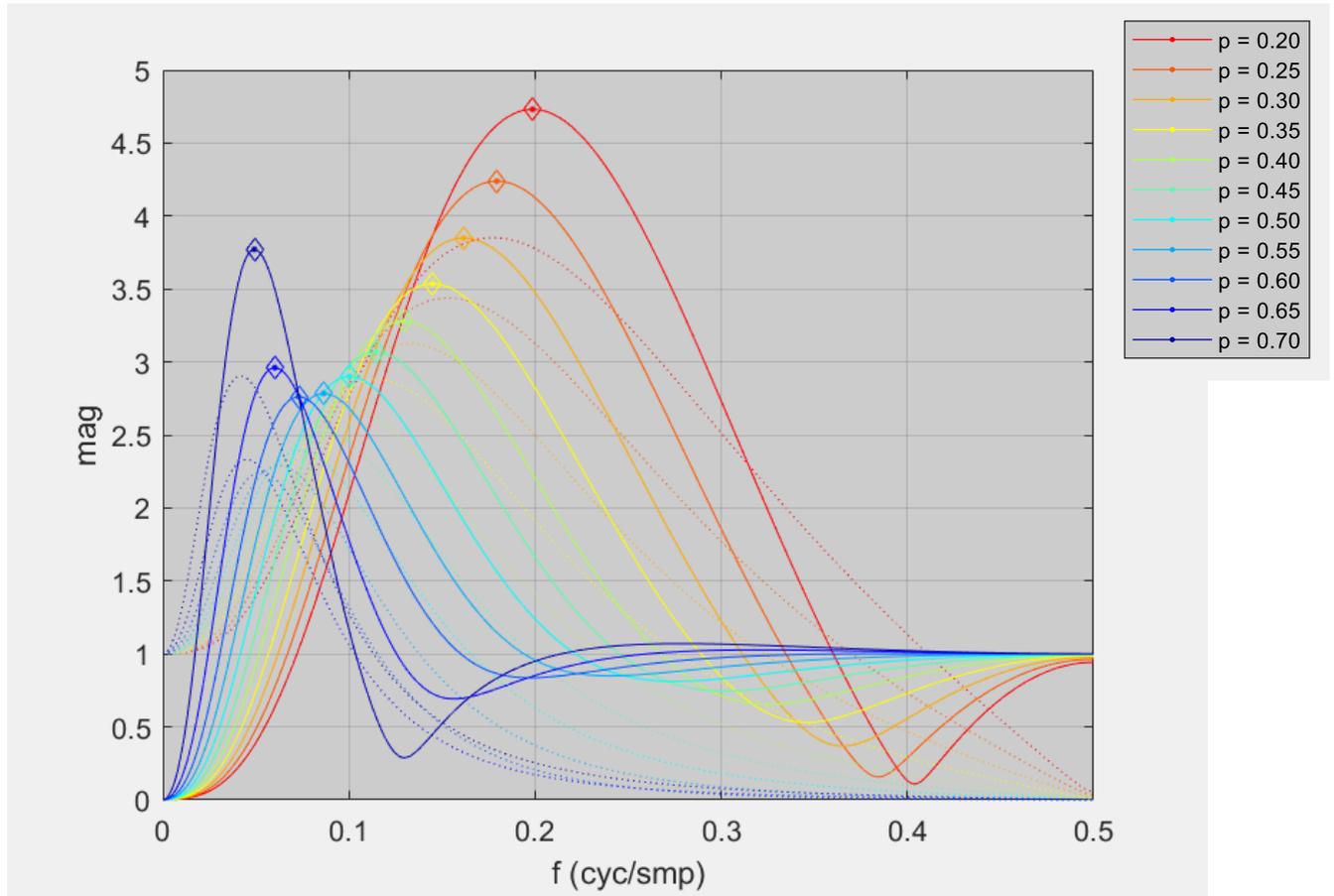

*Figure 11. Fourth-order controller designed via the polynomial method using a variety of different closed-loop pole positions (see legend). Magnitude of sensitivity function (solid lines), sensitivity maxima (diamond tokens) and complementary sensitivity function (dotted lines).*

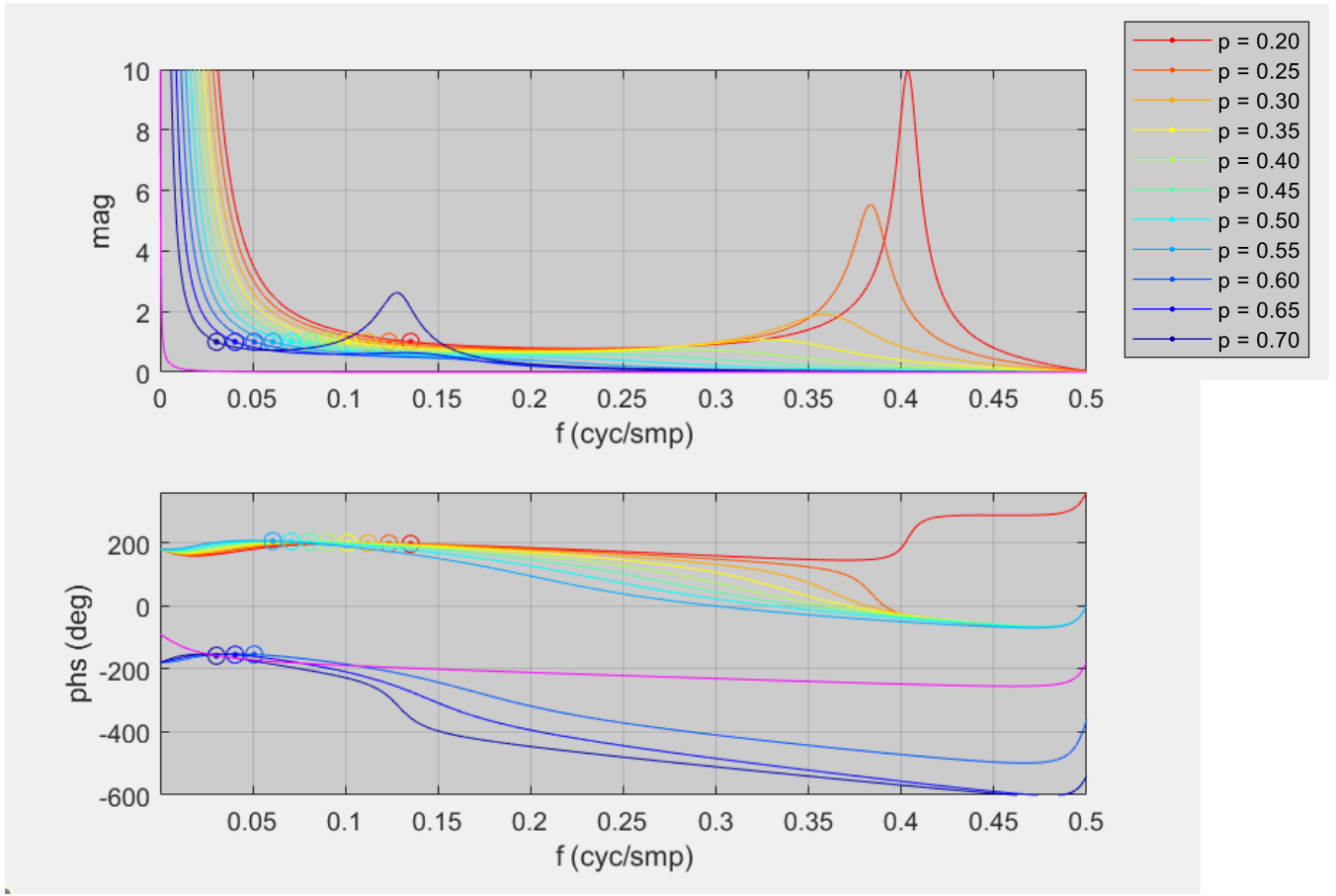

*Figure 12. Fourth-order controller designed via the polynomial method using a variety of different closed-loop pole positions (see legend). Frequency response of loop function (prismatic lines) and plant model (magenta line); magnitude (upper subplot) and phase (lower subplot). Magnitude and phase at gain cross-over frequency of loop function are also shown (circle tokens).*

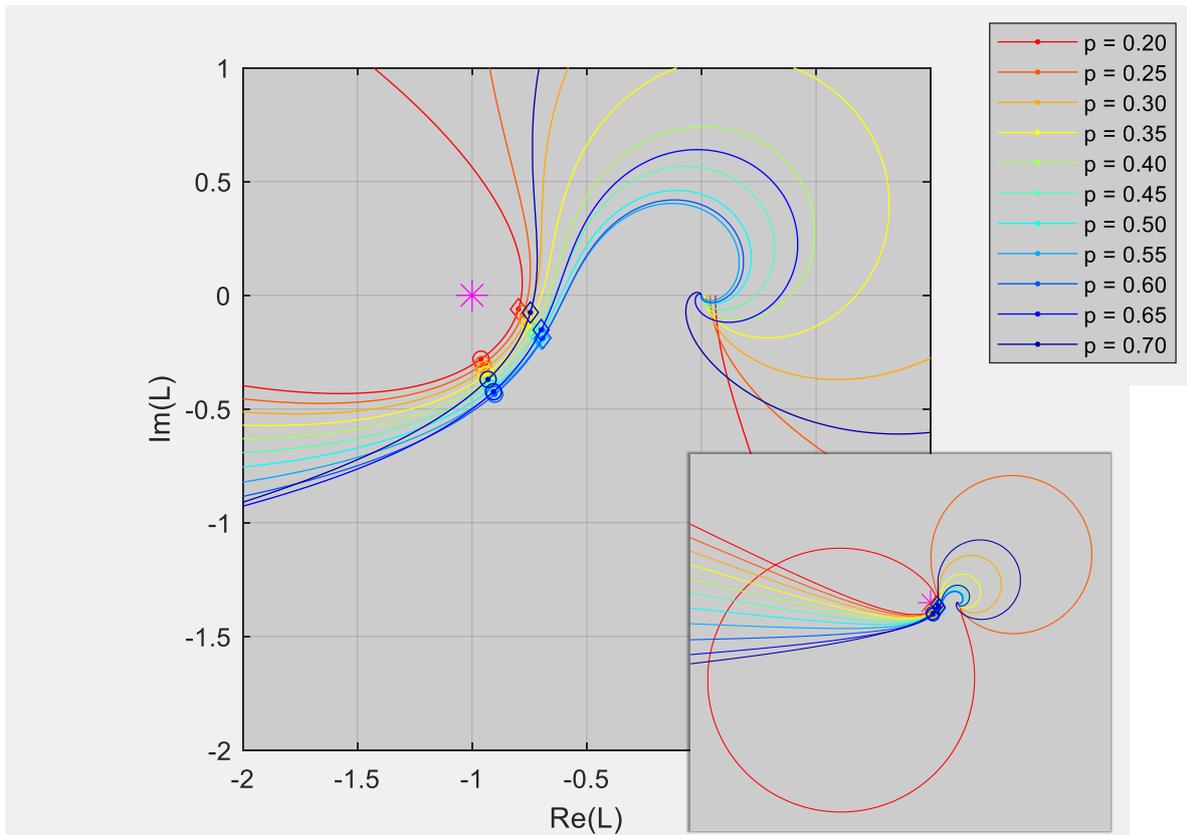

*Figure 13. Fourth-order controller designed via the polynomial method using a variety of different closed-loop pole positions (see legend). Nyquist plots for loop function with phase and magnitude at the gain cross-over frequency (circle tokens) and at the maximum sensitivity frequency (diamond tokens). The critical point at which the magnitude is unity and the phase shift is $\pm\pi$ is also shown (magenta asterisk). Inset shows a wider view and an encirclement of the critical point due to the unstable controller for $p = 0.2$.*

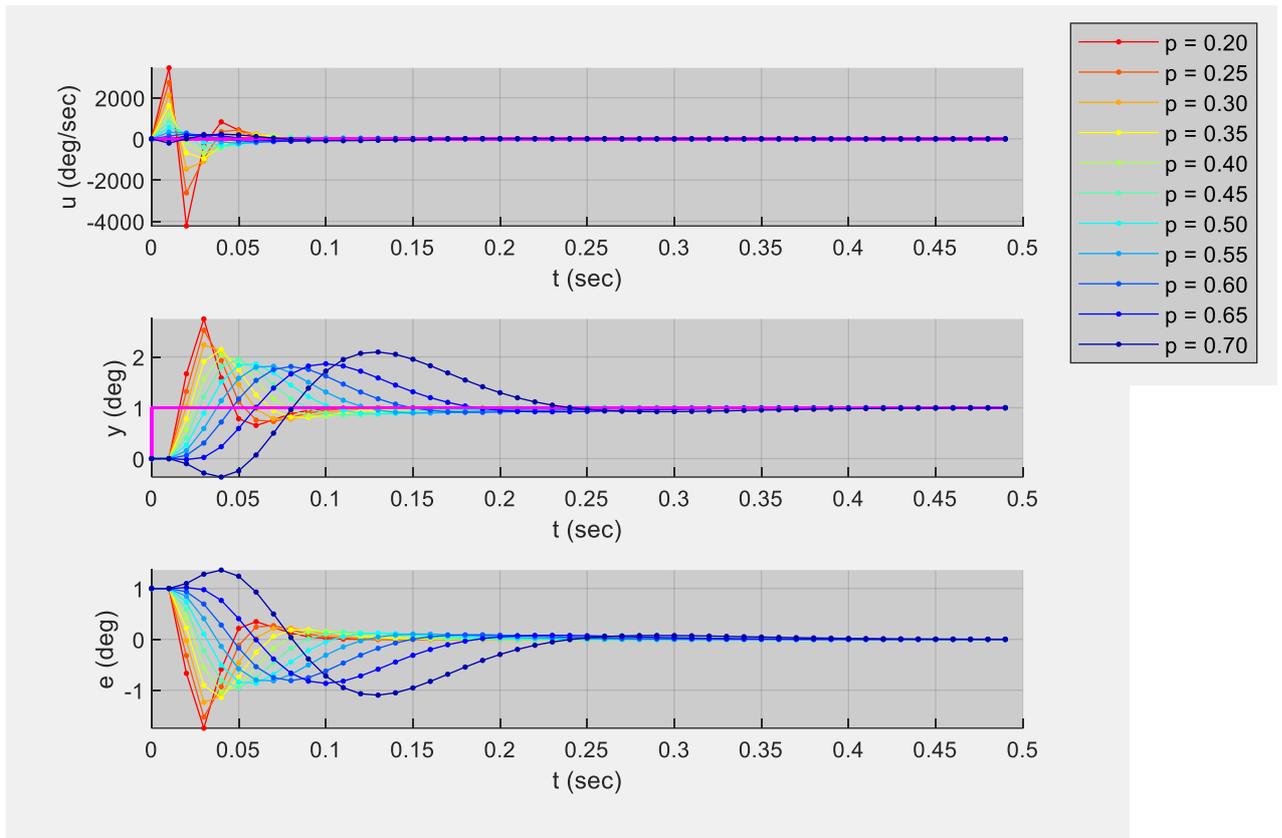

*Figure 14. Fourth-order controller designed via the polynomial method using a variety of different closed-loop pole positions (see legend). Closed-loop response for the nominal plant. Command signal (u, top subplot), plant output (y, middle subplot) and error signal (e, bottom subplot) for a unit step reference (r, magenta line, in middle subplot) and zero disturbances ($d = 0$, magenta line, in top subplot).*

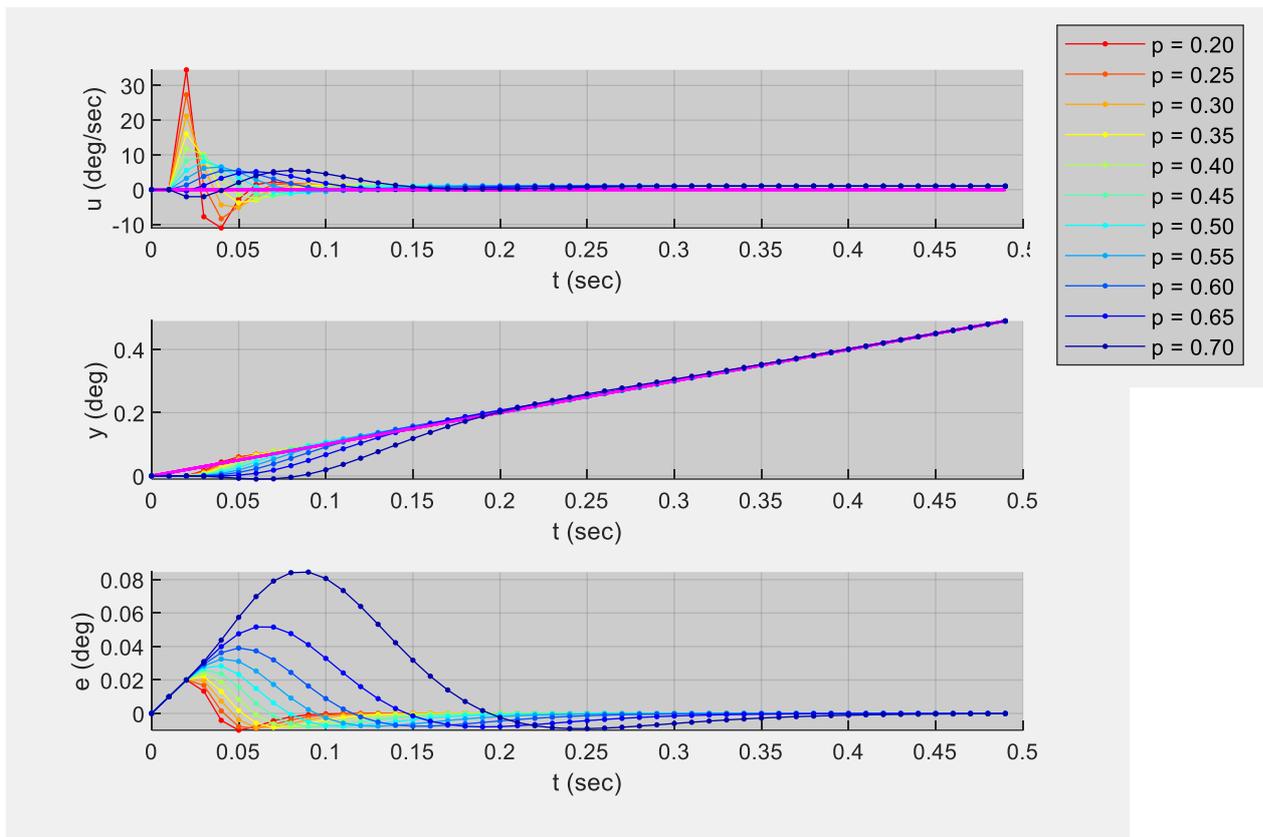

*Figure 15. Fourth-order controller designed via the polynomial method using a variety of different closed-loop pole positions (see legend). Closed-loop response for the nominal plant. Command signal (u, top subplot), plant output (y, middle subplot) and error signal (e, bottom subplot) for ramp reference (r, magenta line, in middle subplot) and zero disturbances ($d = 0$, magenta line, in top subplot).*

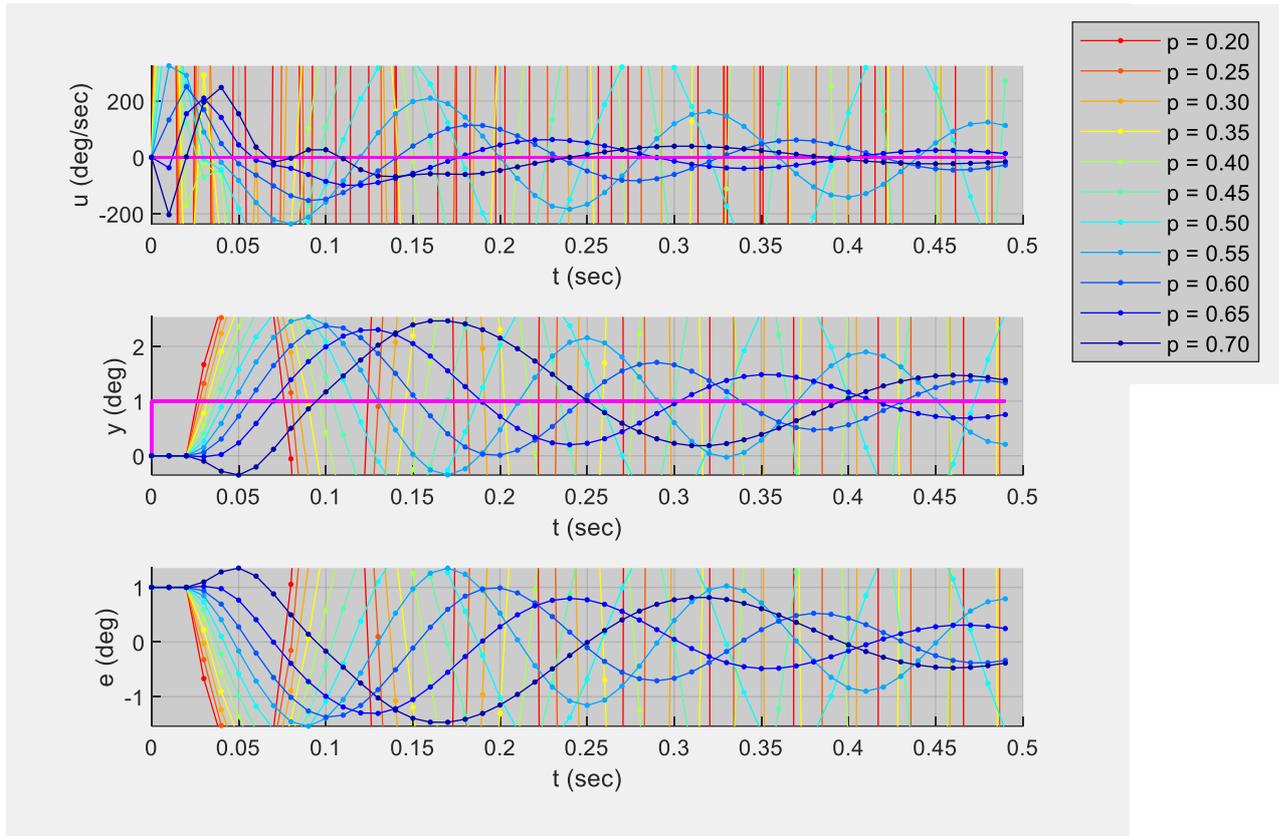

*Figure 16. Fourth-order controller designed via the polynomial method using a variety of different closed-loop pole positions (see legend). Closed-loop response for the perturbed plant with an unmodelled one-sample delay. Command signal (u, top subplot), plant output (y, middle subplot) and error signal (e, bottom subplot) for a unit step reference (r, magenta line, in middle subplot) and zero disturbances ($d = 0$, magenta line, in top subplot).*

*Table 2. Robustness of the fourth-order controller designed via the polynomial method. Phase margin ($\tilde{\varphi}$, deg), delay margin ($\tilde{\Delta}$, smp), and maximum closed-loop pole radius ($r_0$), for the nominal plant model, as a function of closed-loop pole position (p). Maximum closed-loop pole radius for perturbed plant model used in Figure 16 is also shown ($r_1$).*

| $p$ | $\tilde{\varphi}$ | $\tilde{\Delta}$ | $r_0$ | $r_1$ |
|---|---|---|---|---|
| 0.20 | 16.27 | 0.33 | 0.201 | 1.206 |
| 0.25 | 17.66 | 0.40 | 0.251 | 1.168 |
| 0.30 | 19.02 | 0.47 | 0.302 | 1.131 |
| 0.35 | 20.37 | 0.56 | 0.352 | 1.096 |
| 0.40 | 21.69 | 0.66 | 0.402 | 1.064 |
| 0.45 | 22.98 | 0.79 | 0.452 | 1.034 |
| 0.50 | 24.20 | 0.95 | 0.502 | 1.007 |
| 0.55 | 25.24 | 1.15 | 0.552 | 0.984 |
| 0.60 | 25.79 | 1.41 | 0.603 | 0.967 |
| 0.65 | 25.04 | 1.72 | 0.653 | 0.960 |
| 0.70 | 21.69 | 1.98 | 0.703 | 0.964 |

## Analysis of frequency designs

The wider bandwidth tunings for controls designed via the frequency method yield closed-loop poles that are outside the unit circle, e.g. the $\tilde{f} = 0.1$ tuning for the second-order configuration (see Figure 17 & Table 3) and the $\tilde{f} = 0.020$ & $\tilde{f} = 0.018$ tunings for the third-order configuration (see Figure 24 & Table 4). The Nyquist curves and computed stability margins for these systems should be disregarded. The frequency response of these unstable closed-loop systems has no meaning, and the Fourier transform of their impulse responses does not exist, because outputs are unbounded for all non-zero inputs. The frequency-response can, however, be computed from the transfer function, if it is assumed to represent a non-causal system for which the direction of time is reversed. Stable and adroit (i.e. wide bandwidth) controllers cannot be reached for this plant and the PID structure used in the third-order configuration (see Figure 24-Figure 29); however, the stable controllers are very robust (see Figure 30 & Table 4), with respect to unmodelled delays, even though their poles are only just inside the unit circle.

For the second-order configuration, the $\tilde{f} = 0.036$ tuning has the minimum maximum sensitivity (see the diamond tokens in Figure 18 & Figure 20) which indicates that it is the most robust design; however, adroitness (and steady-state tracking) is improved if the bandwidth is widened slightly, e.g. using $\tilde{f} = 0.044$ (see Figure 21 & Figure 22). For the third-order configuration (with an integrator included) decreasing the bandwidth steadily decreases the minimum maximum sensitivity (see Figure 25 & Figure 27) as the Nyquist curve 'flexes' away from the critical point at -1.

*Second-order configuration via frequency design*

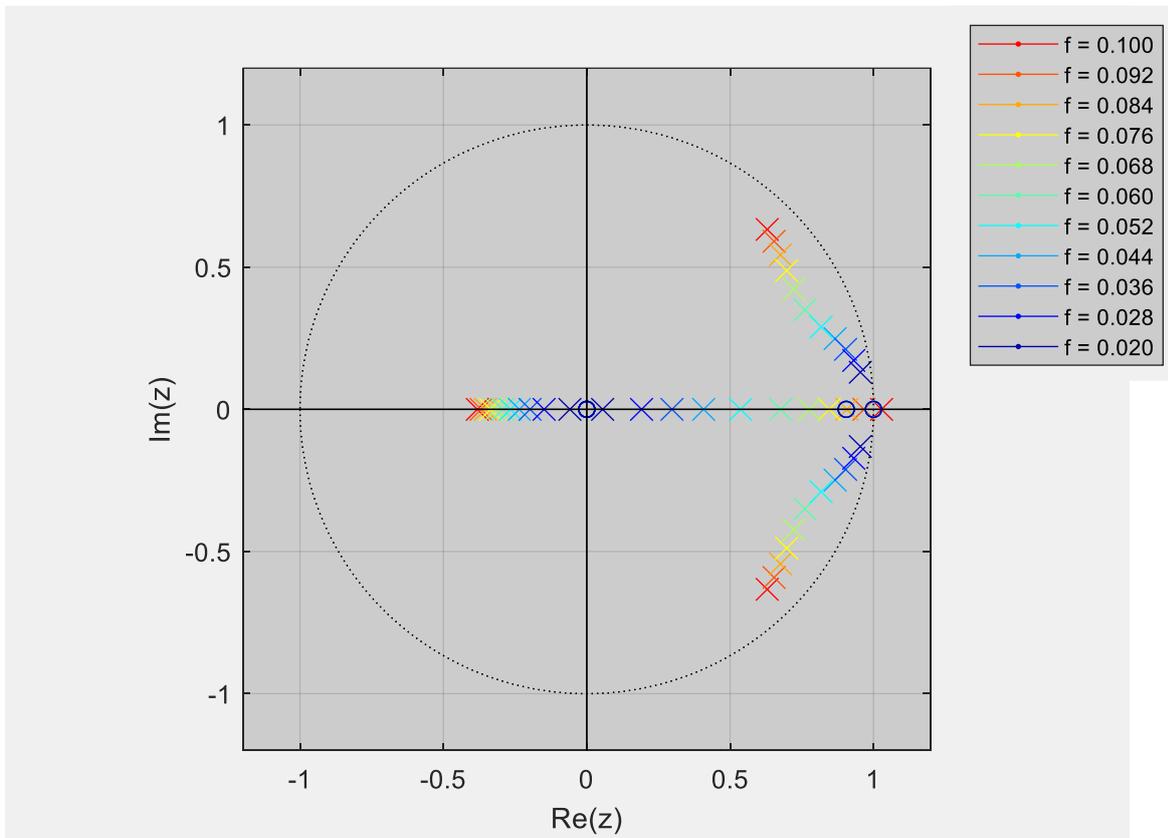

*Figure 17. Second-order controller designed via the frequency method using a variety of different bandwidths (see legend). Positions of the poles ('X' tokens) and zeros ('O' tokens) of the (closed-loop) sensitivity function in the complex z-plane.*

https://arxiv.org/abs/2211.09932

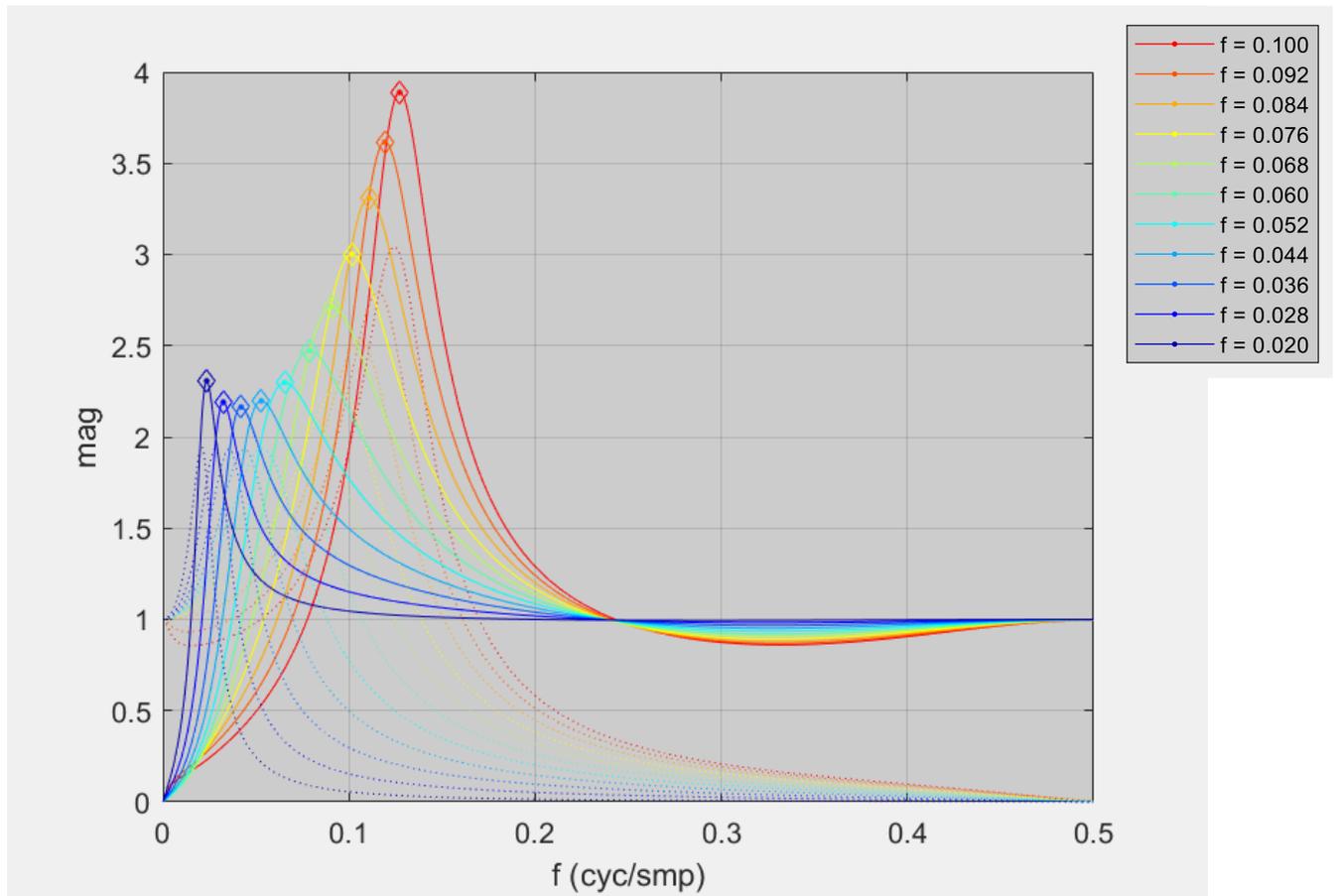

*Figure 18. Second-order controller designed via the frequency method using a variety of different bandwidths (see legend). Magnitude of sensitivity function (solid lines), sensitivity maxima (diamond tokens) and complementary sensitivity function (dotted lines).*

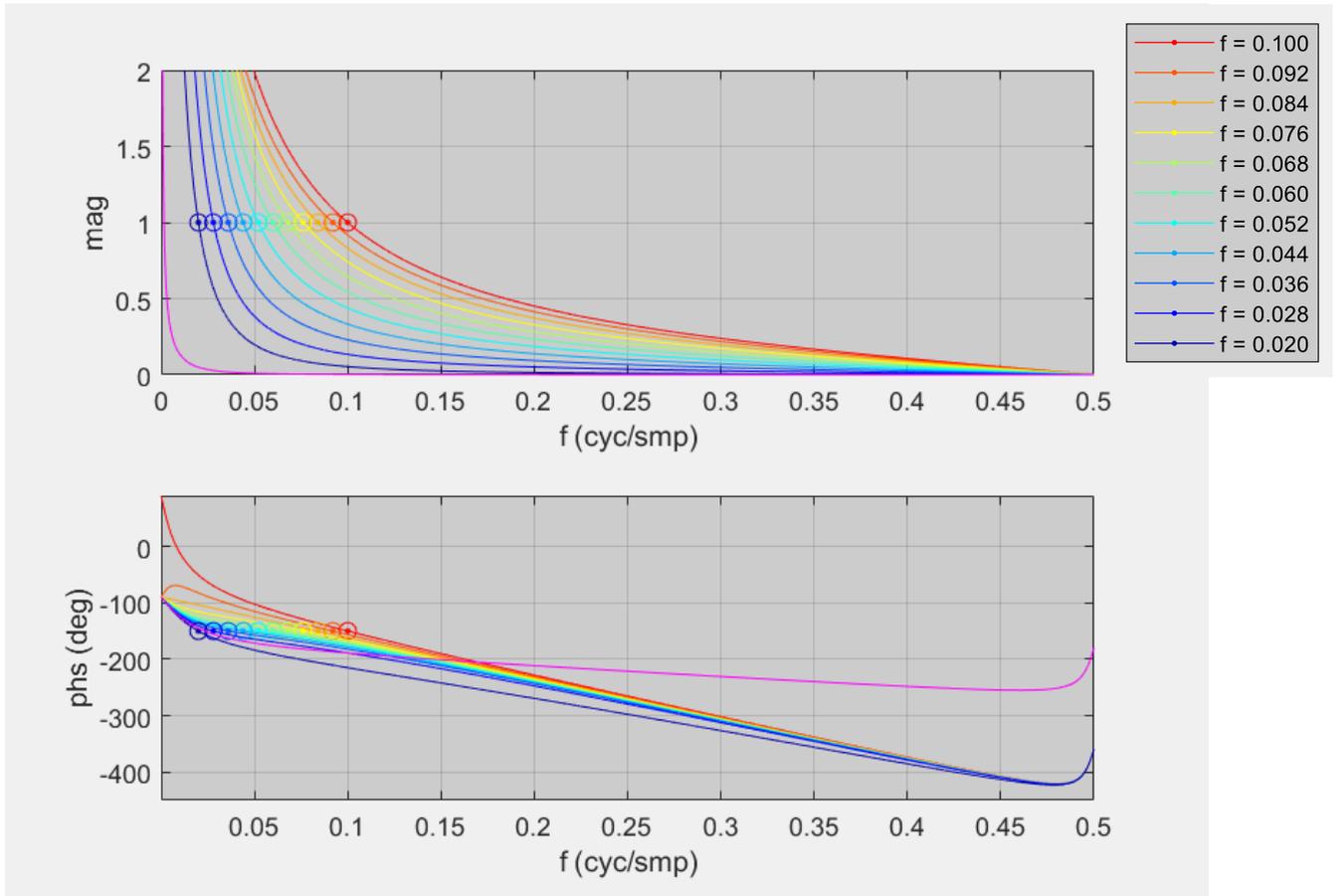

*Figure 19. Second-order controller designed via the frequency method using a variety of different bandwidths (see legend). Frequency response of loop function (prismatic lines) and plant model (magenta line); magnitude (upper subplot) and phase (lower subplot). Magnitude and phase at gain cross-over frequency of loop function are also shown (circle tokens).*

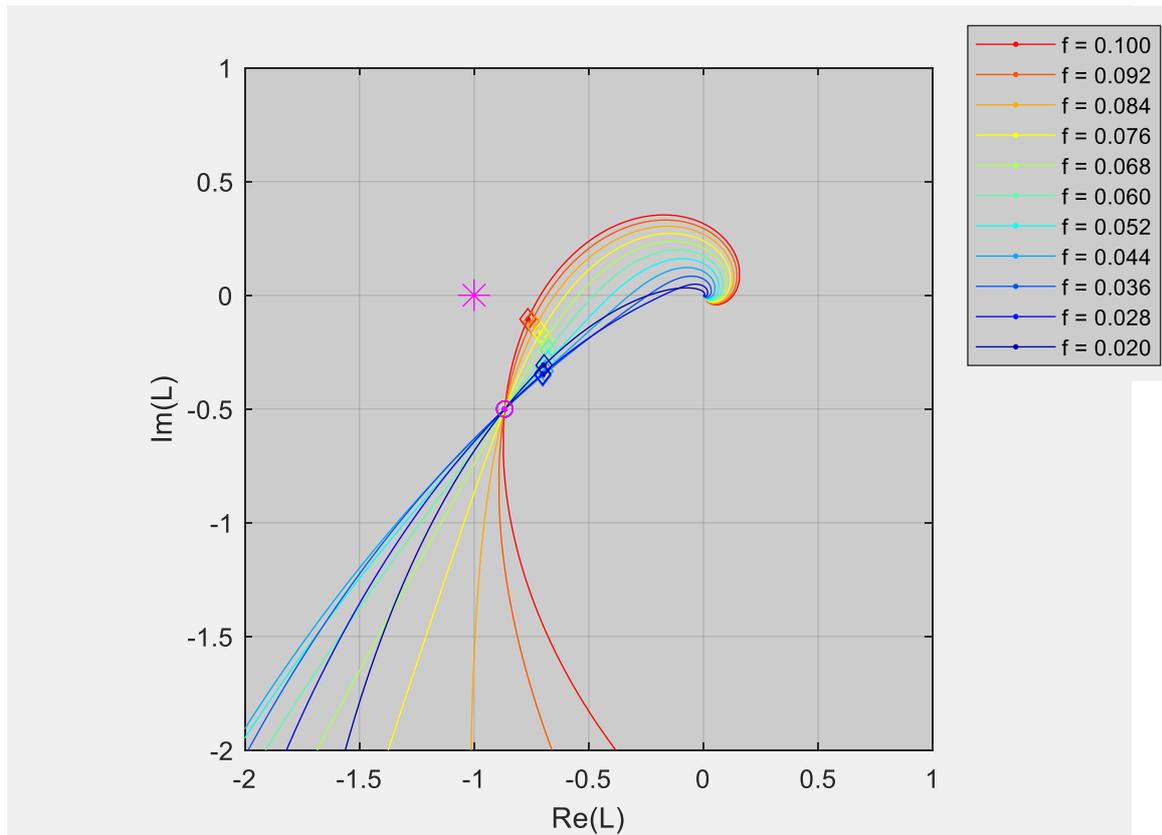

*Figure 20. Second-order controller designed via the frequency method using a variety of different bandwidths (see legend). Nyquist plots for loop function with phase and magnitude at the gain cross-over frequency (magenta circle) and at the maximum sensitivity frequency (diamond tokens). The critical point at which the magnitude is unity and the phase shift is $\pm\pi$ is also shown (magenta asterisk).*

https://arxiv.org/abs/2211.09932

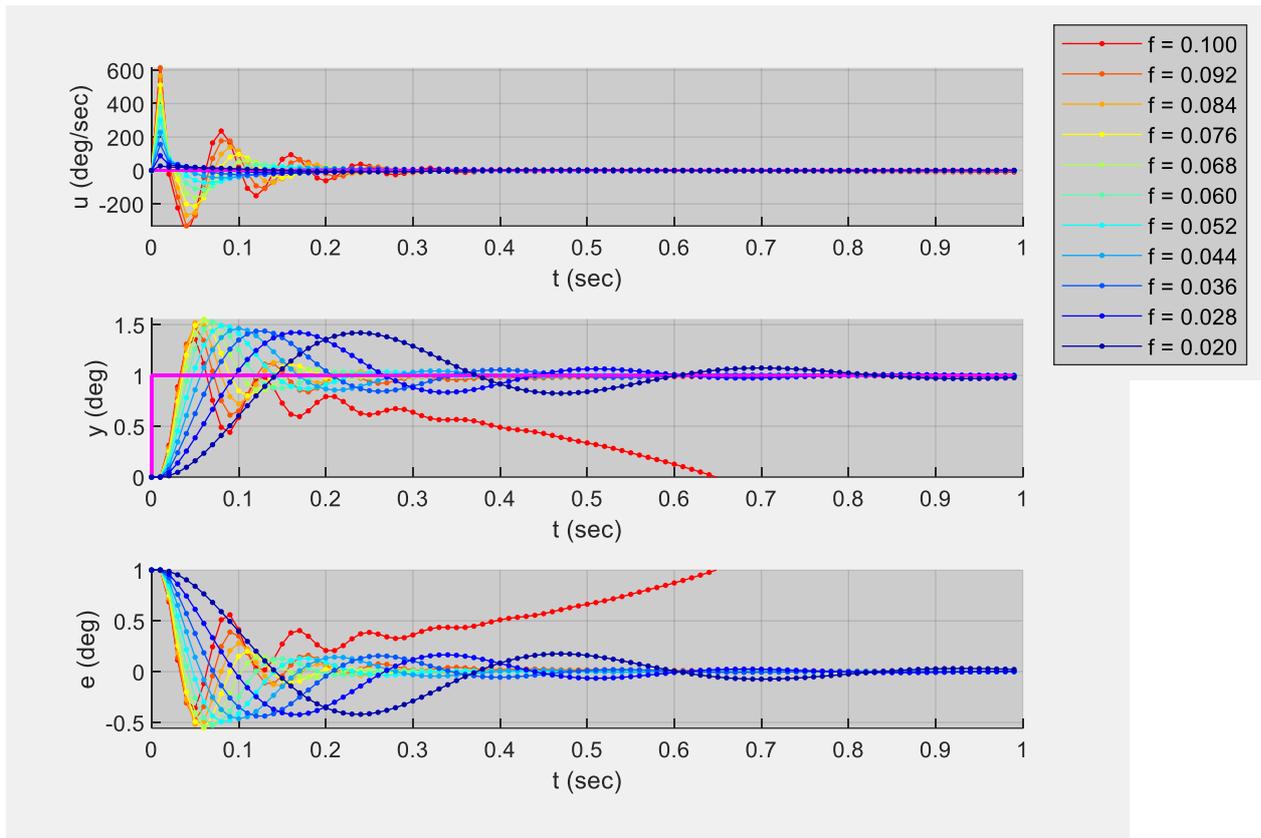

*Figure 21. Second-order controller designed via the frequency method using a variety of different bandwidths (see legend). Closed-loop response for the nominal plant. Command signal (u, top subplot), plant output (y, middle subplot) and error signal (e, bottom subplot) for a unit step reference (r, magenta line, in middle subplot) and zero disturbances (d = 0, magenta line, in top subplot).*

https://arxiv.org/abs/2211.09932

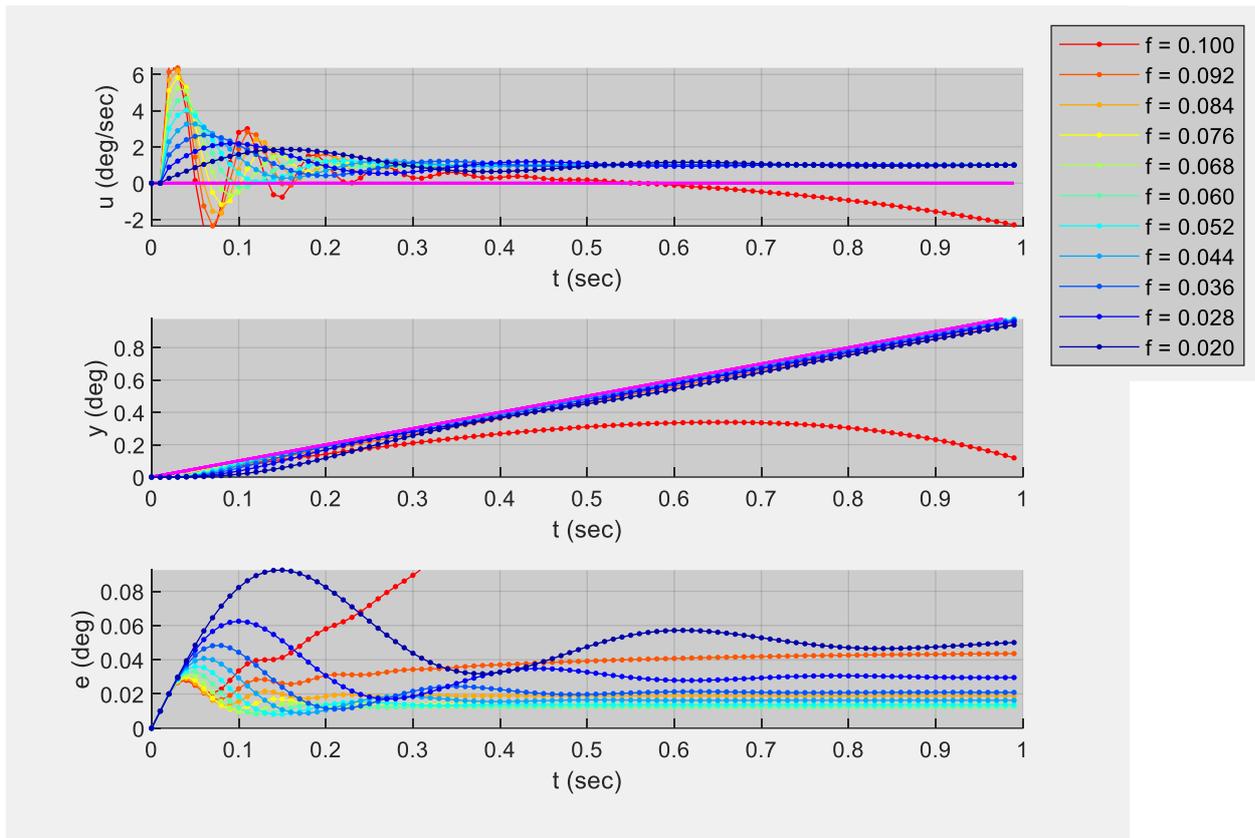

*Figure 22. Second-order controller designed via the frequency method using a variety of different bandwidths (see legend). Closed-loop response for the nominal plant. Command signal (u, top subplot), plant output (y, middle subplot) and error signal (e, bottom subplot) for a ramp reference (r, magenta line, in middle subplot) and zero disturbances (d = 0, magenta line, in top subplot).*

https://arxiv.org/abs/2211.09932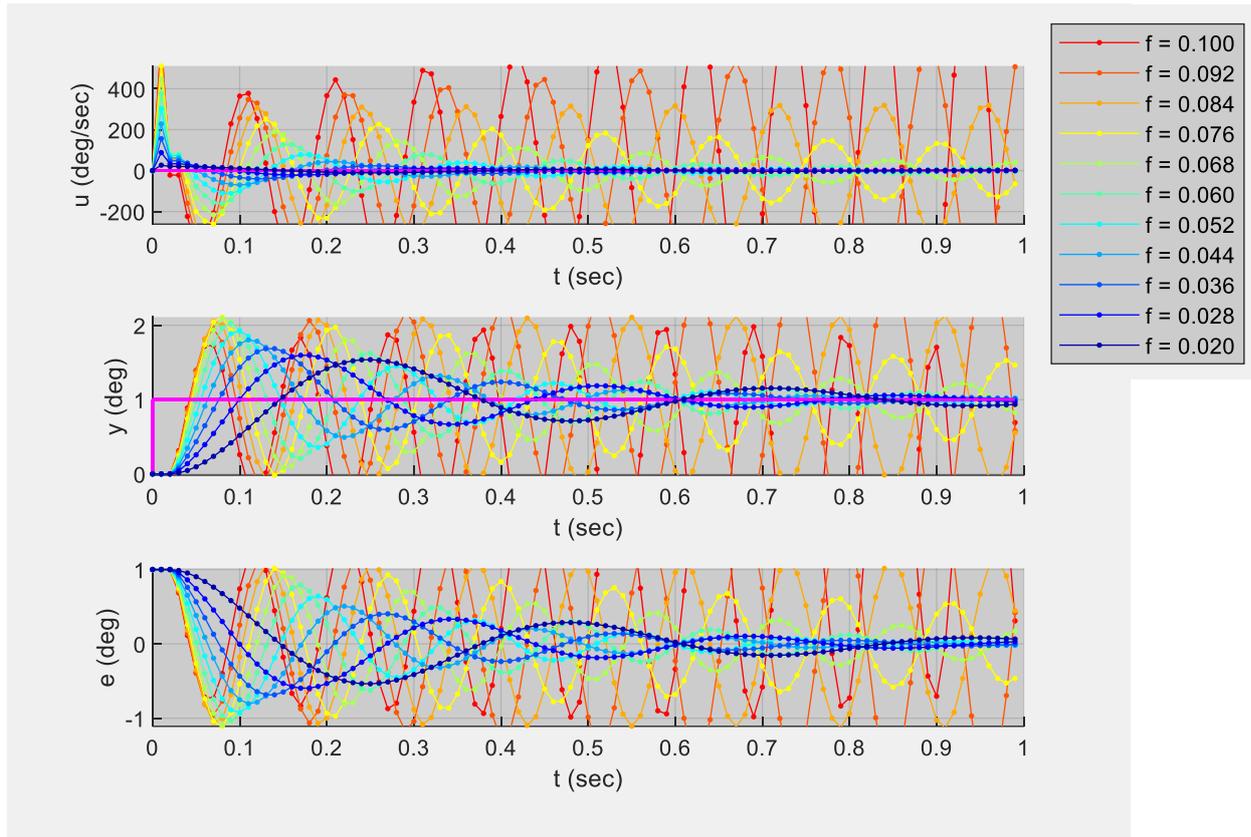

*Figure 23. Second-order controller designed via the frequency method using a variety of different bandwidths (see legend). Closed-loop response for the perturbed plant with an unmodelled one-sample delay. Command signal (u, top subplot), plant output (y, middle subplot) and error signal (e, bottom subplot) for a unit step reference (r, magenta line, in middle subplot) and zero disturbances (d = 0, magenta line, in top subplot).*

*Table 3. Robustness of the second-order controller designed via the frequency method. Phase margin ($\tilde{\varphi}$, deg), delay margin ($\tilde{\Delta}$, smp), and maximum closed-loop pole radius ($r_0$), for the nominal plant model, as a function of controller bandwidth ($\tilde{f}$, cyc/smp). Maximum closed-loop pole radius for perturbed plant model used in Figure 23 is also shown ($r_1$).*

| $\tilde{f}$ | $\tilde{\varphi}$ | $\tilde{\Delta}$ | $r_0$ | $r_1$ |
|---|---|---|---|---|
| 0.100 | 30.00 | 0.83 | 1.0282 | 1.0280 |
| 0.092 | 30.00 | 0.91 | 0.9668 | 1.0073 |
| 0.084 | 30.00 | 0.99 | 0.9071 | 1.0006 |
| 0.076 | 30.00 | 1.10 | 0.8511 | 0.9920 |
| 0.068 | 30.00 | 1.23 | 0.8370 | 0.9814 |
| 0.060 | 30.00 | 1.39 | 0.8373 | 0.9702 |
| 0.052 | 30.00 | 1.60 | 0.8672 | 0.9614 |
| 0.044 | 30.00 | 1.89 | 0.9012 | 0.9581 |
| 0.036 | 30.00 | 2.31 | 0.9271 | 0.9605 |
| 0.028 | 30.00 | 2.98 | 0.9469 | 0.9661 |
| 0.020 | 30.00 | 4.17 | 0.9630 | 0.9733 |

*Third-order configuration via frequency design*

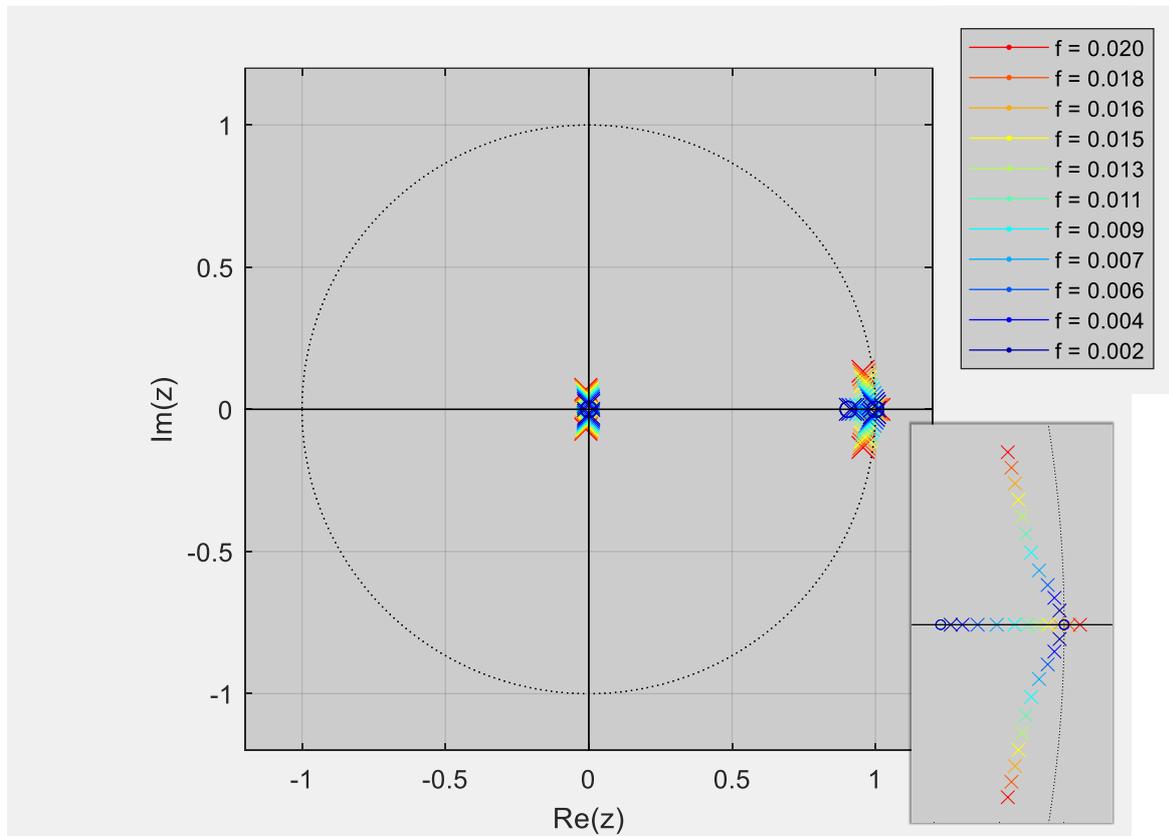

*Figure 24. Third-order controller designed via the frequency method using a variety of different bandwidths (see legend). Positions of the poles ('X' tokens) and zeros ('O' tokens) of the (closed-loop) sensitivity function in the complex z-plane. Inset shows the poles (and zeros) near z = 1.*

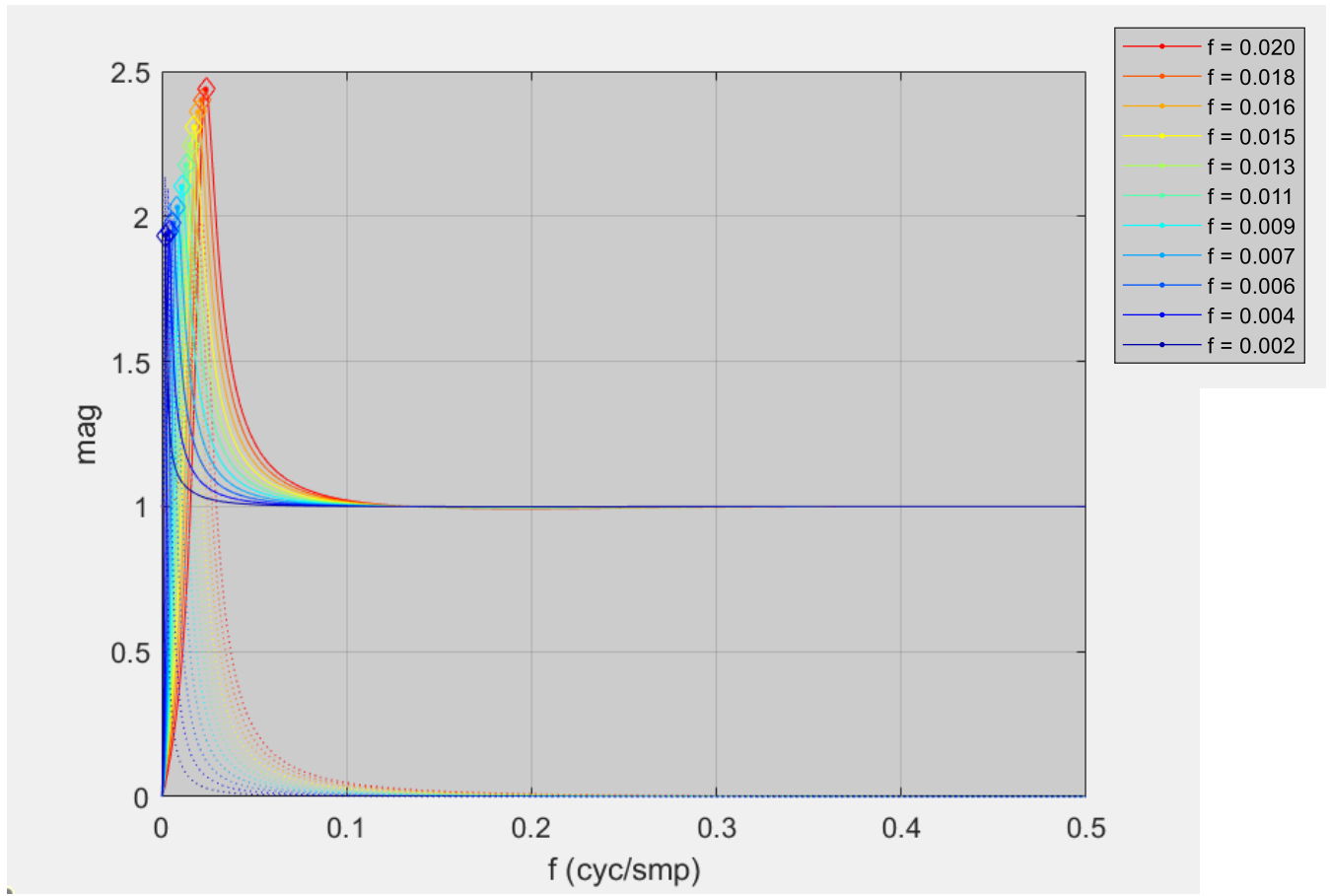

*Figure 25. Third-order controller designed via the frequency method using a variety of different bandwidths (see legend). Magnitude of sensitivity function (solid lines), sensitivity maxima (diamond tokens) and complementary sensitivity function (dotted lines).*

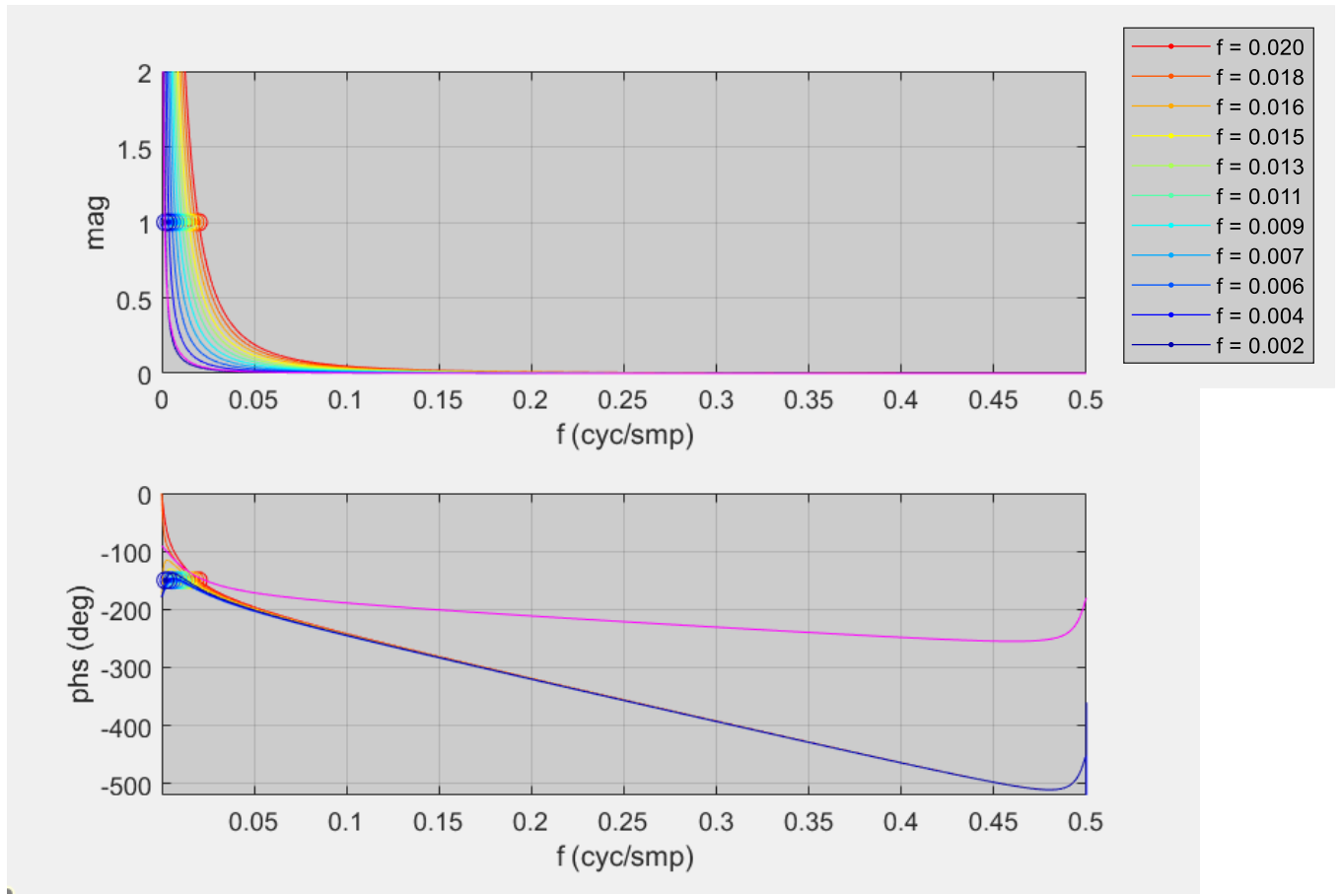

*Figure 26. Third-order controller designed via the frequency method using a variety of different bandwidths (see legend). Frequency response of loop function (prismatic lines) and plant model (magenta line); magnitude (upper subplot) and phase (lower subplot). Magnitude and phase at gain cross-over frequency of loop function are also shown (circle tokens).*

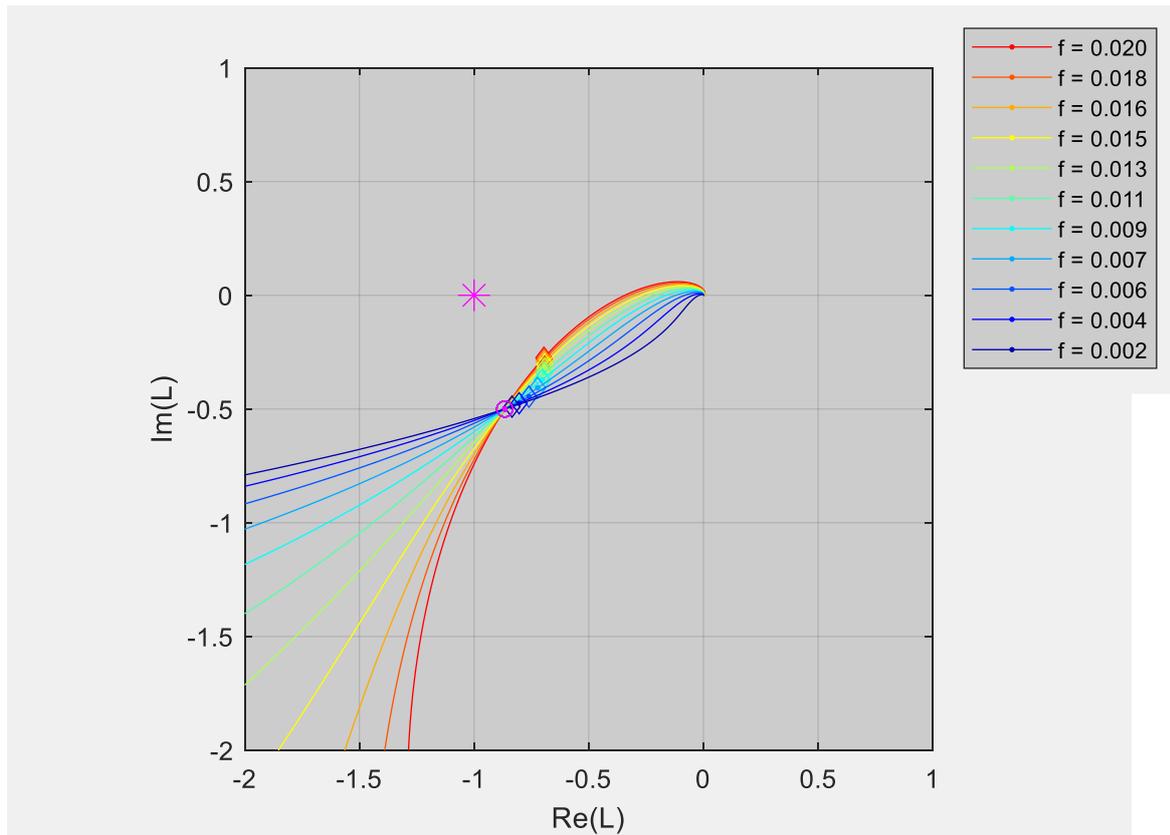

*Figure 27. Third-order controller designed via the frequency method using a variety of different bandwidths (see legend). Nyquist plots for loop function with phase and magnitude at the gain cross-over frequency (magenta circle) and at the maximum sensitivity frequency (diamond tokens). The critical point at which the magnitude is unity and the phase shift is $\pm\pi$ is also shown (magenta asterisk).*

https://arxiv.org/abs/2211.09932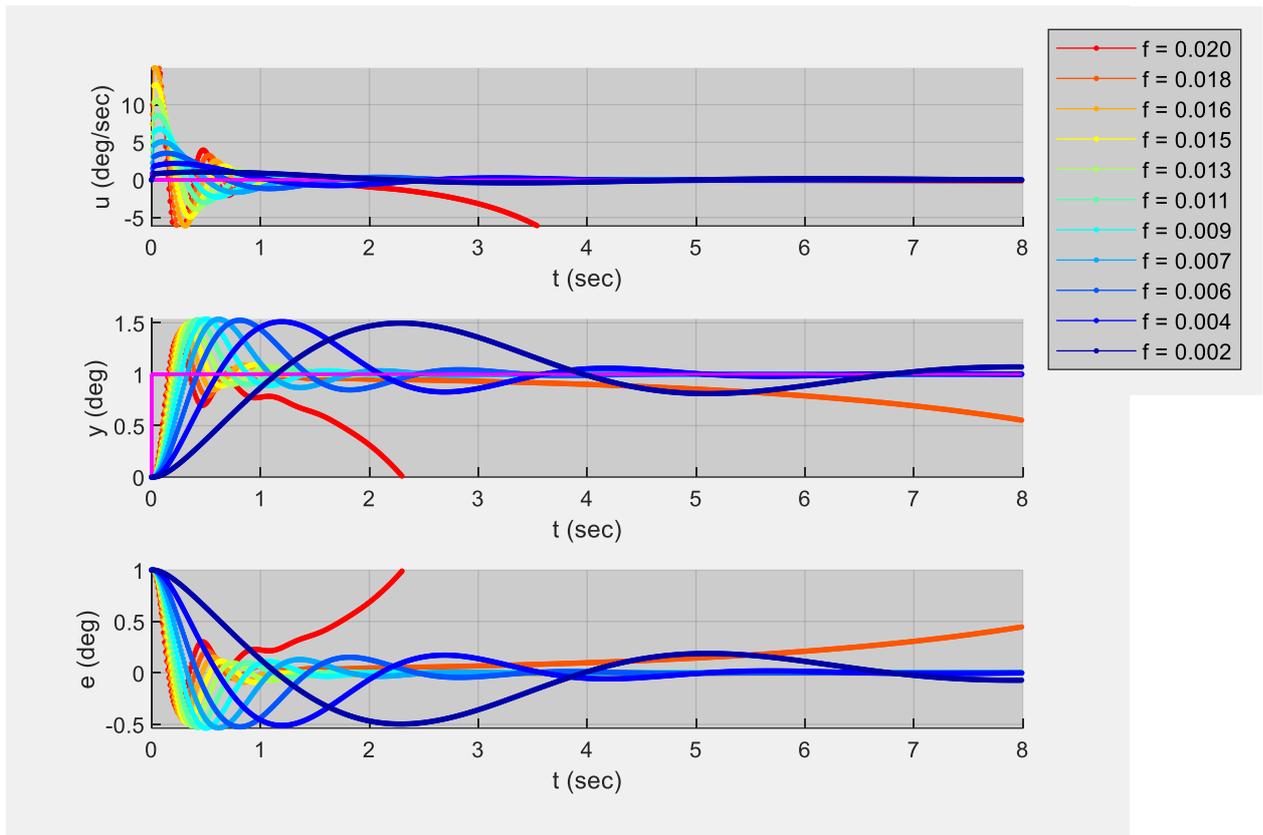

*Figure 28. Third-order controller designed via the frequency method using a variety of different bandwidths (see legend). Closed-loop response for the nominal plant. Command signal (u, top subplot), plant output (y, middle subplot) and error signal (e, bottom subplot) for a unit step reference (r, magenta line, in middle subplot) and zero disturbances ($d = 0$, magenta line, in top subplot).*

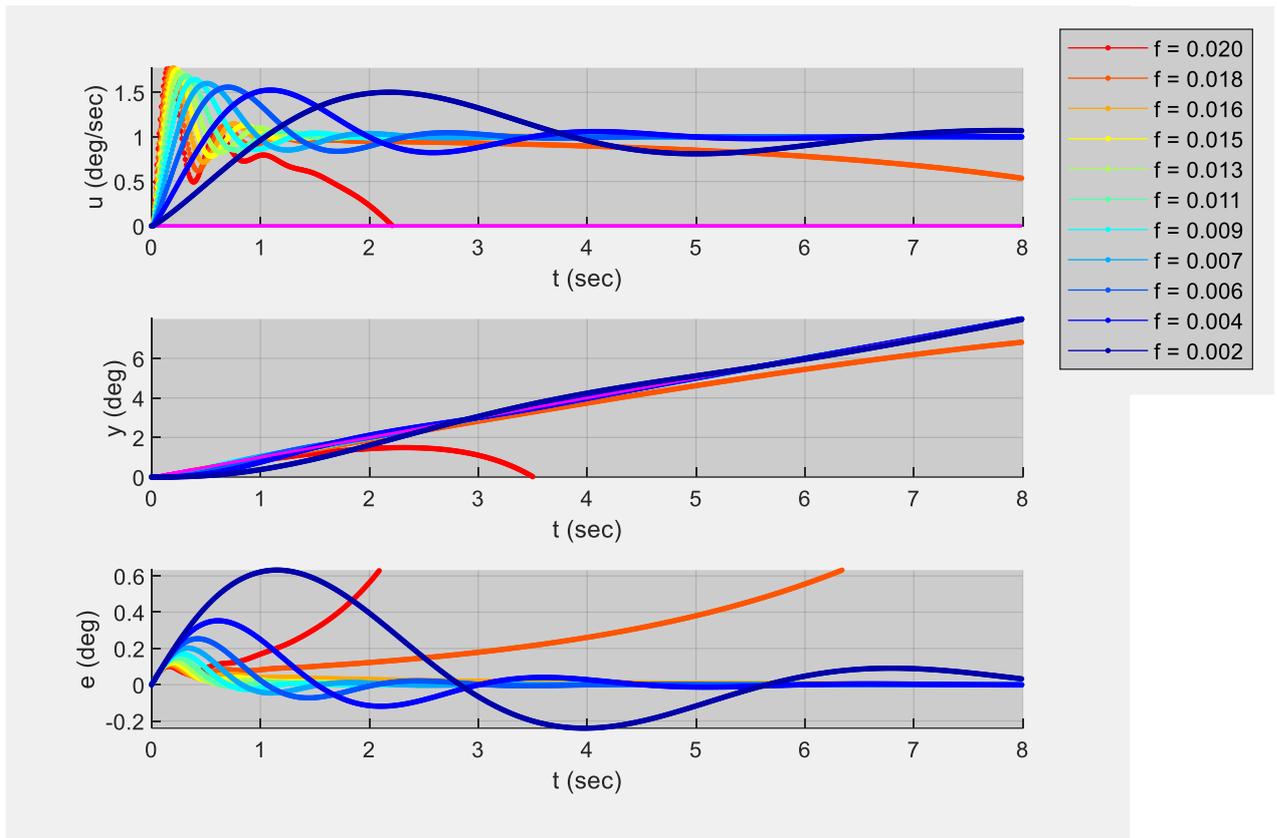

*Figure 29. Third-order controller designed via the frequency method using a variety of different bandwidths (see legend). Closed-loop response for the nominal plant. Command signal (u, top subplot), plant output (y, middle subplot) and error signal (e, bottom subplot) for a ramp reference (r, magenta line, in middle subplot) and zero disturbances (d = 0, magenta line, in top subplot).*

https://arxiv.org/abs/2211.09932

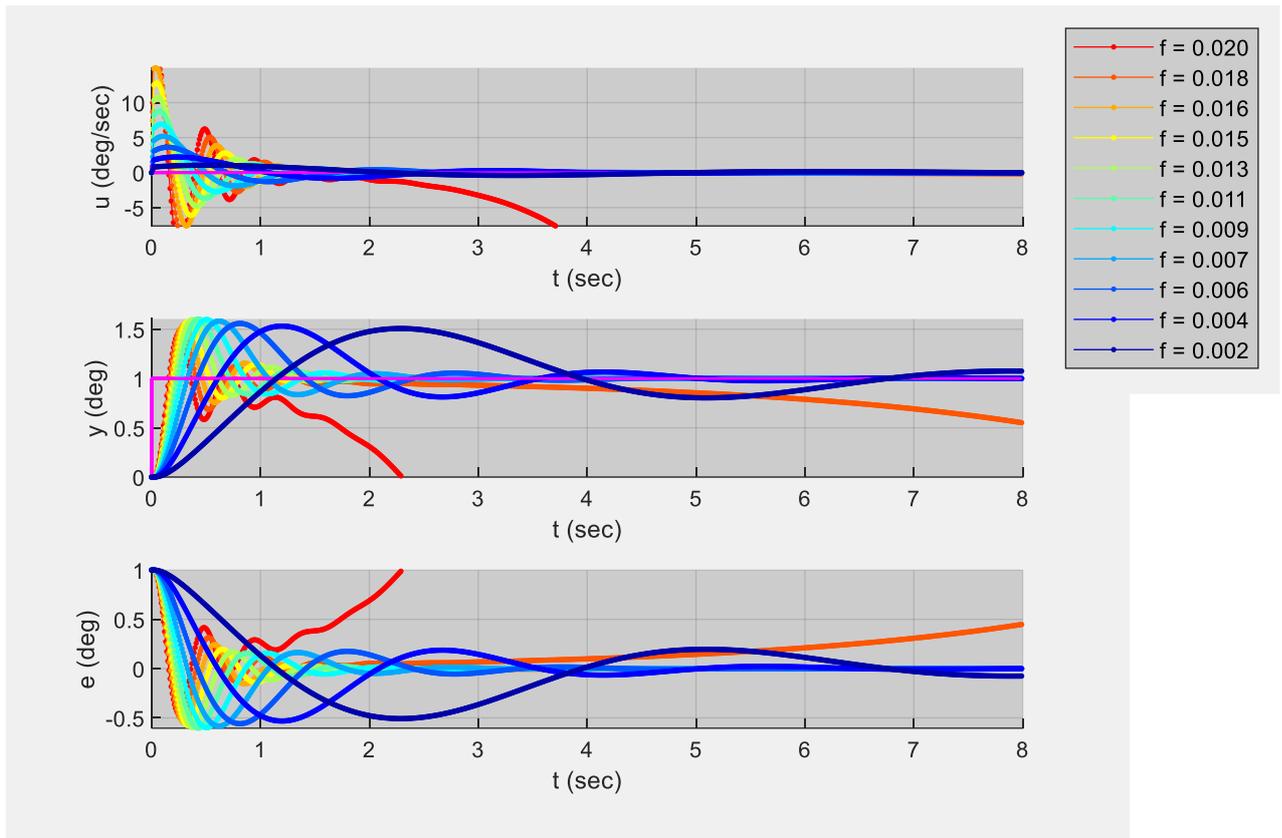

*Figure 30. Third-order controller designed via the frequency method using a variety of different bandwidths (see legend). Closed-loop response for the perturbed plant with an unmodelled one-sample delay. Command signal (u, top subplot), plant output (y, middle subplot) and error signal (e, bottom subplot) for a unit step reference (r, magenta line, in middle subplot) and zero disturbances (d = 0, magenta line, in top subplot).*

*Table 4. Robustness of the third-order controller designed via the frequency method. Phase margin ($\tilde{\varphi}$, deg), delay margin ($\tilde{\Delta}$, smp), and maximum closed-loop pole radius ($r_0$), for the nominal plant model, as a function of controller bandwidth ($\tilde{f}$, cyc/smp). Maximum closed-loop pole radius for perturbed plant model used in Figure 30 is also shown ($r_1$).*

| $\tilde{f}$ | $\tilde{\varphi}$ | $\tilde{\Delta}$ | $r_0$ | $r_1$ |
|---|---|---|---|---|
| 0.020 | 30.00 | 4.17 | 1.0122 | 1.0122 |
| 0.018 | 30.00 | 4.58 | 1.0038 | 1.0038 |
| 0.016 | 30.00 | 5.08 | 0.9957 | 0.9957 |
| 0.015 | 30.00 | 5.71 | 0.9878 | 0.9879 |
| 0.013 | 30.00 | 6.51 | 0.9801 | 0.9801 |
| 0.011 | 30.00 | 7.58 | 0.9732 | 0.9773 |
| 0.009 | 30.00 | 9.06 | 0.9763 | 0.9792 |
| 0.007 | 30.00 | 11.26 | 0.9816 | 0.9830 |
| 0.006 | 30.00 | 14.88 | 0.9878 | 0.9883 |
| 0.004 | 30.00 | 21.93 | 0.9928 | 0.9929 |
| 0.002 | 30.00 | 41.67 | 0.9966 | 0.9966 |

## Closing remarks

Adroit and/or robust control of natural systems in continuous time is possible using only first, second, or third order, digital systems in discrete time that require only a handful of delay, multiply and add operations; thus system performance is certainly *not* limited by computing resources. Determination of appropriate multiplier coefficients by trial and error is laborious at best, often expensive (e.g. for live testing of systems) and almost impossible in some cases (e.g. for highly unstable plants). Fortunately, the theory of linear time-invariant signals-and-systems provides a way; however, the mathematics is not easy to understand, available resources on the topic are daunting, and suggested design techniques & applications are diverse. This document describes two powerful techniques, i.e. the polynomial and frequency methods, that are complementary in many respects; moreover, they are used here to elucidate different aspects of linear time-invariant signals-and-systems theory that are important in aerospace applications; for example in automatic, tuning, stabilising, pointing, steering, homing, and guidance, systems [10].

## Suggested reading and references

[1] P. Kraniauskas, *Transforms in signals and systems*, Wokingham, Addison-Wesley, 1993.
[2] B. P. Lathi, *Linear Systems and Signals*, 2$^{nd}$ Ed, New York, Oxford University, 2005.
[3] Katsuhiko Ogata, *Discrete-Time Control Systems*, 2$^{nd}$ Ed., Prentice Hall, New Jersey, 1995.
[4] K. M. Moudgalya, *Digital Control*, John Wiley & Sons, West Sussex, England, 2007.
[5] M. Grimble, *Robust industrial control systems: optimal design approach for polynomial systems*, Hoboken, New Jersey, Wiley, 2006.
[6] J.C. Doyle, B.A. Francis and A.R. Tannenbaum, *Feedback Control Theory*, Dover Publications, Mineola, N.Y., 2009.
[7] K. Astrom and T. Hagglund, *PID Controllers: Theory, Design, and Tuning*, 2nd Ed., Instrum. Soc. Am., Research Triangle Park, NC., 1995.
[8] H. L. Kennedy, "Optimal design of digital anticipatory servomechanisms with good noise immunity", in Proc. *IEEE Conference on Norbert Wiener in the 21st Century (21CW)*, pp. 1-6, 2016.
[9] G. Stein, "Respect the unstable", *IEEE Control Systems Magazine*, vol. 23, no. 4, pp. 12-25, Aug. 2003.
[10] D. B. Ridgely and M. B. McFarland, "Tailoring theory to practice in tactical missile control", *IEEE Control Systems Magazine*, vol. 19, no. 6, pp. 49-55, Dec. 1999.